\documentclass{article}
\usepackage{authblk}
\usepackage[utf8]{inputenc}
\usepackage[left=2cm,right=2cm,top=1.5cm,bottom=1.5cm]{geometry}
\usepackage{amsmath}
\usepackage{amssymb}
\usepackage{graphicx}
\usepackage{subfig}
\usepackage{mathtools}
\usepackage{mathpazo}
\usepackage[mathpazo]{flexisym}
\usepackage{breqn}
\usepackage{verbatim}
\usepackage{xcolor}
\usepackage{ulem}
\usepackage{placeins}
\usepackage{hyperref}
\usepackage{soul}
\hypersetup{
    colorlinks=true,
    linkcolor=blue,
    filecolor=magenta,      
    urlcolor=cyan,
    citecolor=teal
    }
\usepackage{comment}
\graphicspath{{figs/}}

\usepackage[style=phys,url=false]{biblatex}
\DeclareFieldFormat{titlecase}{#1}
\DeclareFieldFormat{title}{#1}
\DeclareFieldFormat{citetitle}{#1}
\appto\bibsetup{\emergencystretch=3em}

\DeclareMathOperator{\sech}{sech}

\title{Front propagation in non-homogeneous $\phi^4$ model}
\author[1]{Jacek Gatlik}
\author[2]{Tomasz Dobrowolski}
\author[3]{Dominika Lasa}
\author[4]{Panayotis G. Kevrekidis}
\affil[1]{\textit{AGH University of Krakow, Faculty of Physics and Applied Computer Science, 30-059 Krakow, Poland}}
\affil[2]{\textit{Department of Physics and Applied Mathematics, University of the National Education Commission, Podchor\c{a}\.zych  2, 30-084 Krakow, Poland}}
\affil[3]{\textit{Institute of Security and Computer Science, University of the National Education Commission, Podchor\c{a}\.zych  2, 30-084 Krakow, Poland}}
\affil[4]{\textit{Department of Mathematics and Statistics, University of Massachusetts, Amherst, Massachusetts 01003-4515, USA}}

\date{\today}

\begin{document}
\maketitle

\begin{abstract}
We investigate the propagation of fronts in an inhomogeneous medium within the framework of the $\phi^4$ model. The inhomogeneity is modeled either as an interface separating regions with different dissipation or as a finite layer with modified dissipation. The propagating front is described in two ways: as a kink solution in an effectively unbounded domain, and as a half-kink in a finite system. The half-kink represents the decay of the unstable state $\phi=0$ toward the true vacuum. We show that while the effective description based on the kink provides accurate results, applying a similar approach to the half-kink leads to significant deviations from the predictions of the field model. We then demonstrate that a consistent description does exist and propose a modified effective model which reproduces the field-theory results over a relatively broad range of parameters.
\end{abstract} \hspace{10pt}

\section{Introduction}
Nonlinear scalar-field theories provide a compact framework for describing coherent structures and front-like patterns in spatially extended media. A key outcome is the emergence of solitons-robust localized excitations whose dynamics can dominate transport in perturbed environments \cite{Kivshar1989}. In particular, topological solitons represent domain walls interpolating between distinct equilibria, and they arise across condensed-matter settings, including conducting polymers and related low-dimensional materials, as well as broader nonequilibrium pattern-forming contexts \cite{Su1979a,Heeger1988,Cross1993}. Recent studies provide further clarification of universal aspects of domain-wall formation and coarsening dynamics in such nonequilibrium settings \cite{Ma2025}. Similar defect-like structures also play a central role in cosmological settings, where symmetry breaking can generate extended topological defects and domain walls, motivating sustained interest in their dynamics and interactions \cite{Kibble1976,Vilenkin1985,Hindmarsh1995}. It is relevant to highlight that the ever expanding realm of applications of such field theoretic settings has recently included applications in ultracold atomic systems, whereby, e.g., phase dynamics in spinor condensates have been recognized as featuring sine-Gordon or/and double sine-Gordon dynamical evolution~\cite{Yu2021,Siovitz2025}.

Among Klein--Gordon-type models, the $\phi^4$ theory with a double-well potential is a prototypical setting for domain-wall dynamics and front propagation~\cite{KevrekidisCuevasMaraver2019}. Its kink solutions provide a canonical description of interfaces separating stable phases, while front motion toward a preferred vacuum is closely connected to broader themes of interface kinetics and relaxation in bistable media \cite{Allen1979}, consistent also with recent studies of metastable domain-wall configurations in classical field models \cite{Shi2024}. More generally, propagation into unstable or metastable states is a well-developed subject with wide applicability, featuring characteristic selection mechanisms and long-time relaxation of front velocities \cite{Vansaarloos2003,Ebert2000}. Related ideas also appear in vacuum decay problems, where an energetically favored state invades a higher-energy or unstable configuration \cite{Coleman1977}. In realistic media, however, the parameters controlling front motion are rarely uniform, and spatial inhomogeneities can qualitatively reshape dynamics through effective barriers, pinning regions, or transient acceleration/deceleration effects \cite{Kivshar1989,Vansaarloos2003}.

A systematic, time-honored way to interpret the relevant coherent structure dynamics is to reduce the underlying field evolution to a small number of collective coordinates, typically including the front position and, when needed, a width variable \cite{McLaughlin1978a,Keener1977,Rice1983}. This ``effective-particle'' viewpoint has been used extensively to build transparent low-dimensional models for perturbed solitons and to connect quantitative predictions with simulations or experiments \cite{Kivshar1989,McLaughlin1978a,Keener1977,Rice1983,Kevrekidis2004} (see also~\cite{Torrejon2016} for application of collective coordinates in a related magnetic wall setting). In $\phi^4$ specifically, shape dynamics matters because the kink supports internal excitations that can exchange energy with translational motion and noticeably affect scattering and relaxation phenomena \cite{Sugiyama1979,Campbell1983a,Goodman2005,Manton2021}. At the same time, the accuracy and even consistency of collective-coordinate closures can depend sensitively on the ansatz, the chosen degrees of freedom, and the matching of initial conditions, as emphasized in detailed studies of $\phi^4$ kink dynamics and collisions~\cite{Goodman2005}. These efforts are seeking a definitive description qualitatively and especially quantitatively ---as recent publications attest~\cite{Manton2021,Adam2022}---. Additional complications can arise from nontrivial interaction tails and effective forces, which may alter how reduced descriptions capture long-range behavior \cite{Christov2019}.

Spatial heterogeneities and localized defects provide a sharp testbed for such reduced models. Even in the $\phi^4$ setting, localized inhomogeneities can lead to resonant exchange mechanisms, trapping, or reflection, often mediated by interactions between translational motion and internal/defect modes \cite{Fei1992}. More broadly, spatially inhomogeneous media can amplify the role of internal degrees of freedom and lead to length-scale competition effects that are missed by overly rigid reductions \cite{Gonzalez2007}. Related asymmetries and trapping/transmission thresholds have also been studied in defect settings involving balanced gain and loss, highlighting how local energy exchange can strongly modulate kink mobility and phase shifts \cite{Saadatmand2015}. In many physical contexts, dissipation is an unavoidable ingredient; yet it breaks time-reversal symmetry and makes standard conservative variational treatments insufficient. A nonconservative extension of the variational principle provides a consistent route to derive reduced equations of motion with dissipation directly at the action level \cite{Galley2013,Kevrekidis2014}.

It is worth emphasizing that, in its conventional form, the $\phi^4$ model describes an idealized, conservative system, neglecting two ubiquitous features of real physical media: energy dissipation and spatial inhomogeneity. In realistic materials, particularly in condensed matter physics, dissipative processes are unavoidable due to coupling with microscopic degrees of freedom, impurities, phonons, or electronic environments. Moreover, real systems are rarely homogeneous: interfaces, defects, grain boundaries, and engineered heterostructures introduce spatial variations in material properties. As a result, the assumption of a uniform, non-dissipative medium severely limits the applicability of the standard $\phi^4$ framework to experimentally relevant situations. Introducing a spatially dependent dissipation term \cite{Safeer2022,Akosa2016,Jue2016}, constitutes a minimal yet fundamentally important extension of the model. This modification preserves the analytical tractability and conceptual clarity of the $\phi^4$ theory, while incorporating essential physical realism. Moreover it is conducted in the spirit of studies devoted to open systems \cite{Saadatmand2018,Saadatmand2015}. 

In this work, we study front propagation in the $\phi^4$ model in the presence of spatially varying dissipation. The inhomogeneity is modeled either as a smooth interface between two media with different damping or as a finite layer with modified damping. We consider two complementary front scenarios. The first is a driven kink propagating in an effectively unbounded domain, which requires external forcing in the presence of dissipation. The second is a half-kink in a finite domain, representing the decay of the unstable state $\phi=0$ toward the true vacuum. We show that a collective-coordinate approach based on the kink yields an accurate effective description across the heterogeneous region, while a direct analogue for the half-kink can lead to substantial quantitative deviations from the full field evolution. We then demonstrate that a consistent reduced description can be recovered by modifying the effective modeling strategy and introducing an adapted one-coordinate closure that captures the field-theoretic velocity evolution over a broad parameter range. This perspective extends and complements recent work on effective descriptions of kink propagation in heterogeneous or otherwise perturbed settings \cite{Gatlik2023a,Caputo2024,Dobrowolski2025}.

The remainder of the paper is organized as follows. We first define the field model, the inhomogeneous dissipation profiles, and the initial/boundary conditions used for kinks and half-kinks. We then develop an effective description for the driven kink and compare it systematically with the field dynamics for interface and layer configurations. Next, we construct and test standard effective reductions for the half-kink and identify the origins of their mismatch with the field model predictions. Finally, we propose and validate a modified effective model for the half-kink and summarize the resulting conclusions and implications, offering, within our outlook, a number of directions for future study.

\section{Model description}
In this paper, we examine the classical $\phi^4$ model, incorporating a modification that describes media with different damping coefficients
\begin{equation}
\label{phi4+}
\partial_t^2\phi + \eta(x) \partial_t \phi - \partial^2_x\phi + \lambda \, \phi (\phi^2 -1)  = -\Gamma,
\end{equation}
where the function $\eta(x)$ captures the dissipative behavior inherent to the system, specifying how dissipation varies across different regions. 
The external driving term $\Gamma \neq 0$ is necessary to sustain the motion of the kink in a medium with nonzero dissipation. On the other hand, the motion of the half-kink is naturally sustained by the decay of the $\phi=0$ state, and therefore in this case 
---i.e., for the study of the half-kink--- we set $\Gamma=0$.
The constant $\lambda$ describes the field coupling. In this study, we focus on particular choices of the function $\eta(x)$. The first choice corresponds to two media characterized by different values of the dissipation coefficient, with a continuous transition between them
\begin{equation}
\label{eta_function0} \eta(x)= \eta_0 + \frac{1}{2} \varepsilon \left( \tanh(x)+1 \right). 
\end{equation}
Here, $\eta_0$ stands for the reference value of the dissipation, whereas $\varepsilon$ denotes its maximum change, i.e., from
$\eta_0$ to $\eta_0 + \varepsilon$. On the other hand, we also consider the function $\eta$, which describes a sudden change in dissipation localized within a small spatial region, after which it returns to its initial value
\begin{equation}
\label{eta_function} \eta(x)= \eta_0 + \frac{1}{2} \varepsilon \left( \tanh(x)-\tanh(x-\mathtt{L}_0)\right). 
\end{equation}
We denote the width of the region with altered dissipation by $\mathtt{L}_0.$ In the latter case, this corresponds to a localized dissipative layer.
Alternatively, one can think of the former setting of dissipation variation  
as a ``dissipation step'', and of the latter as a ``dissipation box''.

\subsection{Initial conditions for kink configuration}
First, we consider the motion of the kink in a medium with spatially varying dissipation. For this reason, we adopt the following initial conditions
\begin{equation}\label{phi_wp1-kink}
\phi(0,x)= \tanh \left(\sqrt{\frac{\lambda}{2}} \, \gamma_0
\, (x- x_0 ) \right) ,
\end{equation}
\begin{equation}\label{phi_wp2-kink}
\partial_t \phi(0,x)= -\sqrt{\frac{\lambda}{2}} ~ v_0 
\gamma_0 \, \sech^2 \left(  \sqrt{\frac{\lambda}{2}} \, \gamma_0 \, (x-  
x_0) \right) .
\end{equation}
Although $\gamma$ in general describes changes in the width that have both dynamical and kinematical contributions, in the initial condition we assume the value corresponding to the Lorentz factor. The velocity entering the Lorentz factor is taken to be the stationary velocity corresponding to the dissipation present at the initial position 
\begin{equation}
\label{v0-kink}
    v_0 = \frac{1}{\sqrt{1+ \frac{2 \lambda \eta_0^2}{9 \Gamma^2}}} , \,\,\, \gamma_0=\frac{1}{\sqrt{1-v_0^2}} .
\end{equation}
Note that in the case of a kink, an external force is needed to maintain its motion, i.e., $\Gamma \neq 0$. {By analogy with the conditions adopted in \cite{Gatlik2024}, here we also take as the initial value of the velocity the one corresponding to a balance between the external driving force and the dissipation present in the system. The corresponding formula is revisited in what follows in \hyperref[Section3]{Section 3} and obtained below in
Eq.\eqref{v-s-kink+}.} 
{We would like to emphasize that, while we could indeed specify any initial velocity in the initial condition, our focus in this paper is not so much on the kink’s approach to steady-state velocity, but rather on the effect of spatial variations in the dissipation coefficient on the kink’s motion.} 
Additionally, the boundary conditions are chosen to reproduce the asymptotic behavior of the kink, namely $\phi(t,-\ell)=-1$ and $\phi(t,+\ell)=+1$.
\subsection{Initial conditions for half-kink}
Since our second aim is to investigate the decay of the unstable state, we focus on initial conditions representing this scenario: one region of the system is prepared in the true vacuum state, $\phi = 1$, while another is set to $\phi = 0$. As this setup corresponds to a half-kink configuration, we adopt initial conditions consistent with this structure
\begin{equation}\label{phi_wp1-s}
\phi(0,x)= \frac{1}{1+ \exp \left( \sqrt{\frac{\lambda}{2}} \, \gamma_0
\, (x- x_0) \right) },
\end{equation}
\begin{equation}\label{phi_wp2-s}
\partial_t \phi(0,x)= \frac{1}{4} \sqrt{\frac{\lambda}{2}} ~ v_0 
\gamma_0 \, \sech^2 \left(  \frac{1}{2} \sqrt{\frac{\lambda}{2}} \, \gamma_0 \, (x-  
x_0) \right) .
\end{equation}
This form of initial conditions corresponds to an exact solution, in the form of a half-kink, in a system with a constant dissipation coefficient (see \hyperref[AppendixB]{Appendix B}). A major advantage of this choice is that, as we will see in \hyperref[Section5]{Section 5}, 
this waveform features the right asymptotic behavior and thus does not
lead to relaxational velocity oscillations towards the relevant
state.
In reference to the initial conditions we interpret the coefficient $\gamma_0$ as the Lorentz factor. Although, in principle, at initial instant, any propagation velocity could be assigned, we choose a value characteristic to the half-kink solution so as to minimize deviations from its profile at the initial stage of the evolution
\begin{equation}
\label{v0}
    v_0=\frac{1}{\sqrt{1+\frac{2 \eta_0^2}{9 \lambda}}}, \,\,\, \gamma_0=\frac{1}{\sqrt{1-v_0^2}} .
\end{equation}
The system under consideration is confined to the interval $[-\ell, +\ell]$. In the 
case of the half-kink this restriction is motivated not only by the numerical implementation itself, but also by the requirement that the total energy of the configuration remain finite. Such a finite-domain setup naturally aligns with the context of condensed matter physics  where samples are inherently finite.
While finite domains remove the divergence of the energy, it is worth emphasizing that this divergence is absent from the equations of motion themselves. In practice, it is therefore largely harmless, amounting merely to an additive infinite constant in the total energy, and can also be eliminated by an appropriate regularization procedure.
In our simulations, we typically take $\ell = 200$. Furthermore, the boundary conditions are chosen to match the asymptotic behavior of the half-kink, namely $\phi(t,-\ell)=1$ and $\phi(t,+\ell)=0$.

\section{Kink effective model}
\label{Section3}
Our aim herein is to study the motion of the kink in a medium with a position-dependent dissipation coefficient, which simulates the transition from a medium with one dissipation value to a medium with a different dissipation value, as well as the passage through a layer with a modified (smoothly varying) dissipation coefficient
that transitions from one to the other. As a starting point, we consider the case with neither dissipation nor external forcing
\begin{equation}
\label{phi4+k}
\partial_t^2\phi  - \partial^2_x\phi + \lambda \, \phi (\phi^2 -1)  = 0.
\end{equation}
The Lagrangian for the system defined in this way takes the form
\begin{equation}\label{L-k}
L= \int_{-\infty}^{+\infty} dx {\cal L}(\phi)
=\int_{-\infty}^{+\infty} dx  \left[ \frac{1}{2}~ (\partial_t
\phi)^2 - \frac{1}{2}  (\partial_x \phi)^2 - \frac{1}{4} \lambda \left(\phi^2 - 1 \right)^2  \right].
\end{equation}
To begin with, we introduce a new field variable 
consonant with the kink asymptotics that we intend to study through the transformation
\begin{equation}\label{kink-xi-k}
\phi_K(t,x) =  \tanh \xi(t,x).
\end{equation}
The Lagrangian written in terms of the new field variable is as follows
\begin{equation}
\begin{gathered}
\label{Lxi}
L = \int_{-\infty}^{+\infty} dx  \,\, {\rm sech}^4 \xi \left(
\frac{1}{2}\,   (\partial_{t} \xi)^2 
-   \frac{1}{2}\,   (\partial_{x} \xi)^2 
 - \frac{1}{4}~\lambda 
   \right).
\end{gathered}
\end{equation}
The restriction to the sector containing the single kink solution is implemented by adopting an Ansatz that reduces the dynamics to two dynamical degrees of freedom
\begin{equation}
    \label{xi-k}
    \xi(t,x)= \sqrt{\frac{\lambda}{2}}\,  \gamma(t) (x - x_0(t)) .
\end{equation}
The first degree of freedom, $x_0(t)$, describes the position, while the second, $\gamma(t)$, is the inverse of the kink’s width.
After performing integration with respect to the spatial variable, we obtain the effective Lagrangian
\begin{equation}\label{Leff-k}
L_{eff} = \frac{1}{2}\, M \Dot{x}_0^2 +\frac{1}{2}\, m
\Dot{\gamma}^2  - V.
\end{equation}
The mass coefficients in this Lagrangian depend only on the (inverse) width of the kink and are independent of its position 
\begin{equation}
\begin{gathered}
\label{integrals-k}
    M = \frac{1}{2} \lambda \gamma^2 \int_{-\infty}^{+\infty}dx ~\sech^4(\xi)  = \frac{4}{3}  \gamma \sqrt{\frac{\lambda}{2}},\\
    m = \frac{1}{\gamma^2}\int_{-\infty}^{+\infty}dx ~ \sech^4(\xi) \, \xi^2 = \frac{1}{ \gamma^3} \sqrt{\frac{2}{\lambda}} \left( \frac{\pi^2}{9} - \frac{2}{3} \right).
\end{gathered}
\end{equation}
Since the Lagrangian contains no spatial inhomogeneities, the effective potential is also independent of the kink position [reflecting the translational invariance
of the model]
\begin{equation}
\label{v-eff-k}
    V = \frac{1}{4} \lambda\int_{-\infty}^{+\infty}dx \,  \sech^4(\xi) \,  \left(   \gamma^2  + 1 \right) = \frac{2}{3 \gamma}  \sqrt{\frac{\lambda}{2}} \left(   \gamma^2  + 1 \right) .
\end{equation}
To incorporate dissipation into the system, we employ the formalism introduced and developed in Refs. \cite{Galley2013, Kevrekidis2014}.
In its conventional, conservative form, the variational principle is formulated as a boundary-value problem in time, meaning that the dynamical variables, or fields, are specified both at the beginning and at the end of the evolution.  While this formulation is perfectly suited to reversible dynamics, it is fundamentally incompatible with genuinely irreversible phenomena, since it treats forward and backward time evolution symmetrically.  To address this limitation, Refs. \cite{Galley2013, Kevrekidis2014} introduced a modification of the variational framework itself. Rather than fixing the dynamical variables at both temporal endpoints, only the initial state is prescribed, leaving the final-time configuration unconstrained. This adjustment is implemented indirectly through a doubling of the system’s degrees of freedom: each original variable is replaced by a pair of “copies,” each evolving along an independent trajectory. An extended Lagrangian, which depends on both copies, is then constructed so that the variational principle constrains variations at the initial time while allowing them to remain free at the final time. In the final stage, known as taking the physical limit, the two copies are identified, yielding the true physical trajectory of the system. The key advantage of this doubled-variable approach is that it naturally incorporates dissipative effects. By breaking the temporal symmetry inherent in the traditional formulation, it allows for a consistent description of processes where energy is not conserved. Within this non-conservative variational framework, the corresponding Lagrangian density is expressed as:
\begin{equation}
    \label{nonconservative}
    {\cal L}_N = {\cal L}(\phi_1) - {\cal L}(\phi_2) + {\cal R} .
\end{equation}
In the above formula, the first two terms represent the Lagrangian densities for both copies of the field $\phi$, while the last term is responsible for the effects associated with dissipation. In this extended formalism, the equations of motion preserve much of the structural form characteristic of the standard conservative theory. The primary difference is the emergence of an extra term, which directly captures the impact of dissipative processes within the system. This additional contribution alters the conventional dynamics, allowing the evolution to reflect energy dissipation or other non-conservative effects in a consistent manner. In general, for the Lagrangian density ${\cal L}_N$, the field equation takes the form
\begin{equation}
    \label{eq-nonconservative}
    \partial_{\mu} \left( \frac{\partial {\cal L}}{\partial (\partial_{\mu} \phi)}\right) - \frac{\partial {\cal L}}{\partial \phi} = \left[\frac{\partial {\cal R}}{\partial \phi_{-}} - \partial_{\mu} \left( \frac{\partial {\cal R}}{\partial (\partial_{\mu} \phi_{-})}\right)\right]_{PL}.
\end{equation}
Here, the index $\mu$ runs over the space-time coordinates $x^{\mu}=(x^0,x^1)=(t,x)$, and the summation convention is assumed. We also introduce the variables $\phi_{+}$ and $\phi_{-}$, which are connected to the original fields $\phi_1$ and $\phi_2$ through the relations
\begin{equation}
    \phi_1=\phi_{+}+ \frac{1}{2}\phi_{-} \,\,\,\,\,
\mathrm{and} \,\,\,\,\,\phi_2=\phi_{+}- \frac{1}{2}\phi_{-}
\end{equation}
Importantly here, PL denotes the ‘physical limit’, that is, the limit in which the auxiliary fields $\phi_+$ and $\phi_-$ reduce as $\phi_-  \rightarrow 0$ and $\phi_+ \rightarrow \phi$.
In the special case of the $\phi^4$ model, the above equation can be expressed as
\begin{equation}
\label{eq-nonconservative2}
    \partial_t^2 \phi - \partial^2_x  \phi + \lambda \phi (\phi^2-1) = \left[\frac{\partial {\cal R}} {\partial \phi_{-}} - \partial_{\mu} \left( \frac{\partial {\cal R}}{\partial (\partial_{\mu} \phi_{-})}\right)\right]_{PL} .
\end{equation}
By taking the expression for ${\cal R}$, describing dissipation 
(proportional to $\eta(x)$)
and external forcing (proportional to $\Gamma$), in the form 
\begin{equation}
   {\cal R} = - \Gamma \phi_- - \eta(x) \phi_{-} \partial_t \phi_{+} .
\end{equation}
from \eqref{eq-nonconservative2}, we will reproduce equation \eqref{phi4+}.
Notice that in this expression $\eta$ can be a function of $x$, as will be the case
in the present work.
Since the definition of the non-conservative contribution is not unique, in our work we also include external constant forcing in this term of the Lagrangian. After integrating over the variable $x$, we obtain an expression that describes dissipation at the level of the effective model 
\begin{equation}
  R_{eff} = \int_{-\infty}^{+\infty} dx {\cal R} .  
\end{equation}
In general, the equations of the two-degree-of-freedom effective model $(x_0,\gamma)$, incorporating dissipation, take the form
\begin{equation}
    \label{eff-eq1}
    \frac{d}{d t }\left(\frac{\partial L_{eff}}{\partial \Dot{x}_0} \right) - \frac{\partial L_{eff}}{\partial {x}_0} = \left[ \frac{\partial R_{eff}}{\partial {x}_{-}} -
    \frac{d}{d t }\left(\frac{\partial R_{eff}}{\partial \Dot{x}_{-}} \right) \right]_{PL} ,
\end{equation}
\begin{equation}
    \label{eff-eq2}
    \frac{d}{d t }\left(\frac{\partial L_{eff}}{\partial \Dot{\gamma}} \right) - \frac{\partial L_{eff}}{\partial {\gamma}} = \left[ \frac{\partial R_{eff}}{\partial {\gamma}_{-}} -
    \frac{d}{d t }\left(\frac{\partial R_{eff}}{\partial \Dot{\gamma}_{-}} \right) \right]_{PL} .
\end{equation}
In a similar vein as above, PL denotes the ‘physical limit’, that is, the limit in which the auxiliary spatial variables $x_+$ and $x_-$ reduce to $x_-  \rightarrow 0$ and $x_+ \rightarrow x_0$. An analogous characterization arises in the case of
the collective coordinate $\gamma$.
The relations between one set of variables and the other are as follows $x_1=x_+ + \frac{1}{2} x_-$, $x_2=x_+ - \frac{1}{2} x_-$ and $\gamma_1=\gamma_+ + \frac{1}{2} \gamma_-$, $\gamma_2=\gamma_+ - \frac{1}{2} \gamma_-$.
Within the formalism discussed above, we obtain a system of ordinary differential equations (in the physical limit)
\begin{equation}
\begin{gathered}
\label{2dof_ansatz-k}
    M\Ddot{x}_0+ \frac{1 }{\gamma}  M \, \Dot{x}_0 \Dot{\gamma}
     =2 \Gamma -A \Dot{x}_0 - N \Dot{\gamma} ,\\
    m\Ddot{\gamma} +\frac{1}{2}(\partial_{\gamma}m)\Dot{\gamma}^2
    - 
    \frac{1}{2} (\partial_{\gamma}M) \Dot{x}_0^2 
  +   \partial_{\gamma}V=- A_{\gamma} \Dot{x}_0 - N_{\gamma} \Dot{\gamma}.
\end{gathered}
\end{equation}
The coefficients appearing in the equations of motion are determined by integral expressions
\begin{equation}
    \label{wsp}
A= \frac{1}{2} \lambda \gamma^2 \int_{-\infty}^{+\infty} dx \, \eta(x)  \sech^4 \xi , \quad
    N = - \sqrt{\frac{\lambda}{2} }  \int_{-\infty}^{+\infty} dx \, \eta(x) \xi \sech^4 \xi ,
\end{equation}
\begin{equation}
    \label{wsp2}
A_{\gamma}= - \sqrt{\frac{\lambda}{2} }   \int_{-\infty}^{+\infty} dx \, \eta(x)  \xi \sech^4 \xi , \quad
    N_{\gamma} =  \frac{1}{\gamma^2}   \int_{-\infty}^{+\infty} dx \, \eta(x) \xi^2 \sech^4 \xi .
\end{equation}
Here, as can be observed in the above equations, we are incorporating the
dissipation spatial profile that will be examined in what follows.
Note that coefficients $A_{\gamma}$ and $N$ are equal. Based on the above equations, we can determine the steady-state velocity of the kink resulting from the balance between dissipation and the forcing present in the system.
For $\eta=const\equiv \eta_0$, assuming $\Dot{x}_0=v=const$ and $\gamma=const$ from the first equation of the system \eqref{2dof_ansatz-k}, we obtain the steady-state value of the speed
\begin{equation}
\label{v-s-kink+}
    v_0 = \frac{1}{\sqrt{1+ \frac{2 \lambda \eta_0^2}{9 \Gamma^2}}} .
\end{equation}
We begin comparing the results of the field model with the effective model by describing evolution of the dynamical variable $x_0$. 

{Figure~\ref{fig_01}, panel (a), shows a comparison between the trajectory obtained from the field equation \eqref{phi4+} (solid black line) and that from the effective model \eqref{2dof_ansatz-k} (orange dashed line). The gray region represents the time during which the kink is within a region of modified dissipation. The agreement between the two trajectories is indeed very good.
However, a closer inspection of the difference illustrated in panel (b), defined as $\Delta x_0(t) = x_0^{PDE}(t) - x_0^{ODE}(t)$, reveals a subtle beating pattern. Both panels correspond to the case where the initial position of the kink is $-50$, the dissipation coefficient is fixed at $\eta_0 = 0.1$, the driving force is $\Gamma = 0.025$, and the variation in the dissipation coefficient is set to $\varepsilon = 0.05$.
It turns out that a more natural quantity for describing the evolution is the velocity as a function of time. This quantity will be presented in the subsequent figures throughout this work.
\begin{figure}[h!]
    \centering
    \subfloat{{\includegraphics[height=4.5cm]{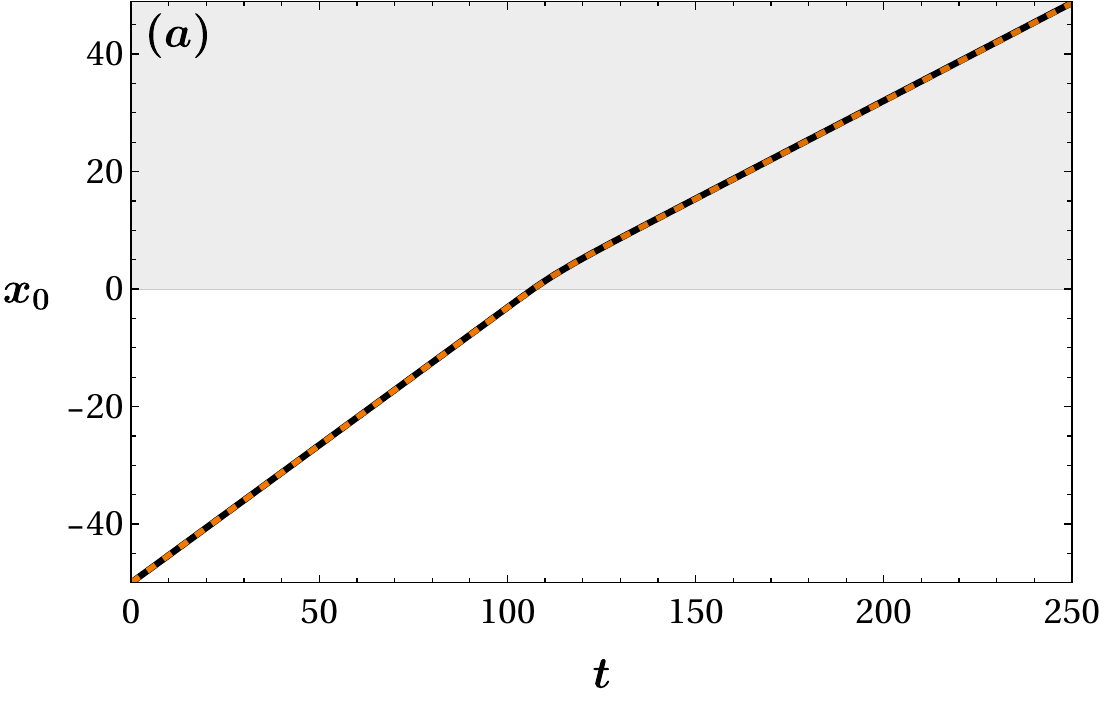}}}
    \quad
    \subfloat{{\includegraphics[height=4.5cm]{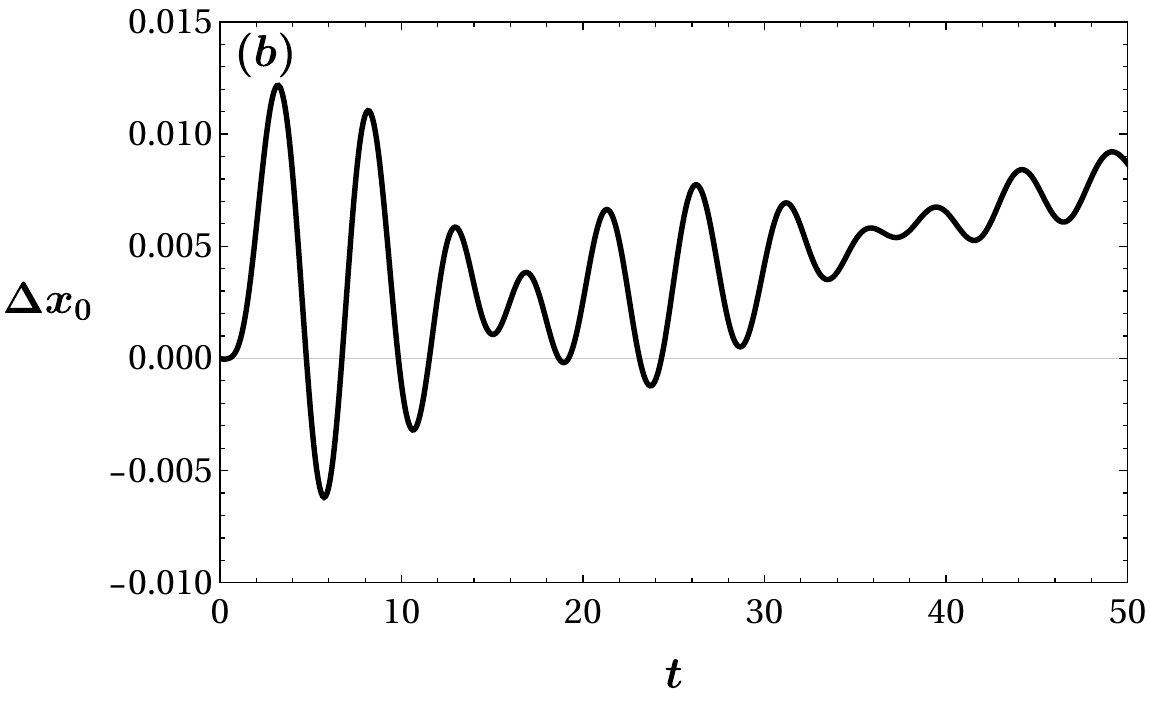}}}
    \caption{{(a) Comparison of kink trajectories from the field equation (solid black line) and the effective model (orange dashed line). The gray area represents the time during which the kink moves in the area of changed dissipation. Panel (b) shows the difference $\Delta x_0(t)$. Parameters are chosen as follows: $x_0=-50$, $\eta_0 = 0.1$, $\Gamma = 0.025$, $\lambda = 1$, and $\varepsilon = 0.05$.}}
    \label{fig_01}
\end{figure}

The changes in the velocity of the kink during the transition from a medium with lower dissipation to one with higher dissipation are illustrated in Figure~\ref{fig_02}.}
The gray area in the figures shows the time when the kink enters the region of higher dissipation.
The initial speed of the kink is set based on the formula \eqref{v-s-kink+}. Panels (a) and (b) show a reduction in kink velocity, which is reproduced quite accurately by the effective model. Note that relative to the value $\eta_0$ in the first medium, dissipation in the second medium changes by fifty (a) and one hundred (b) percent. 
It is important to highlight here that when the kink is introduced in the medium,
there exists a transient stage at the beginning of the motion. This is related to the fact that $\gamma$ in the initial condition is chosen as the Lorentz factor for the stationary velocity $v_0$ and does not take into account dynamical changes in the kink thickness associated with the presence of dissipation and external forcing. 
Accordingly, the kink adjusts its motion early on to accommodate the presence
of the relevant dissipative evolution.

\begin{figure}[h!]
    \centering
    \subfloat{{\includegraphics[height=4.5cm]{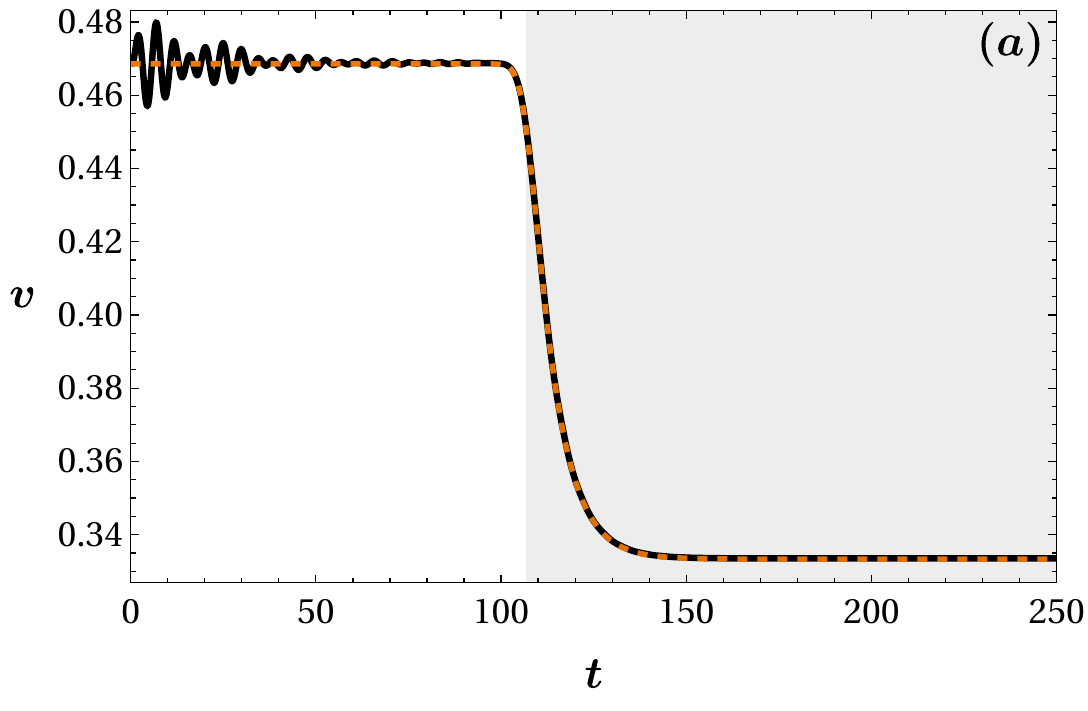}}}
    \quad
    \subfloat{{\includegraphics[height=4.5cm]{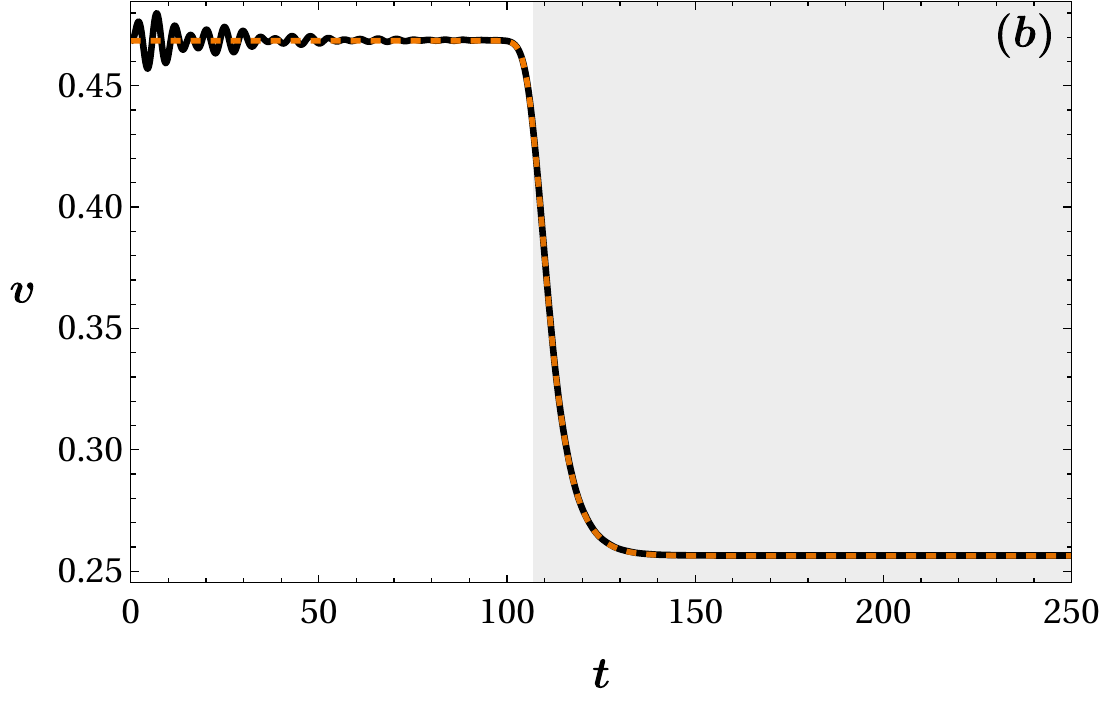}}}
    \caption{Changes in kink velocity when entering an area of increased dissipation. In both panels, the initial position of the kink is $x_0=-50.$ Parameters in the figures are chosen as follows $\eta_0 = 0.1$, $\Gamma = 0.025$, $\lambda = 1$, (a) $\varepsilon = 0.05$, (b) $\varepsilon = 0.1$.}
    \label{fig_02}
\end{figure}

\begin{figure}[h!]
    \centering
    \subfloat{{\includegraphics[height=4.5cm]{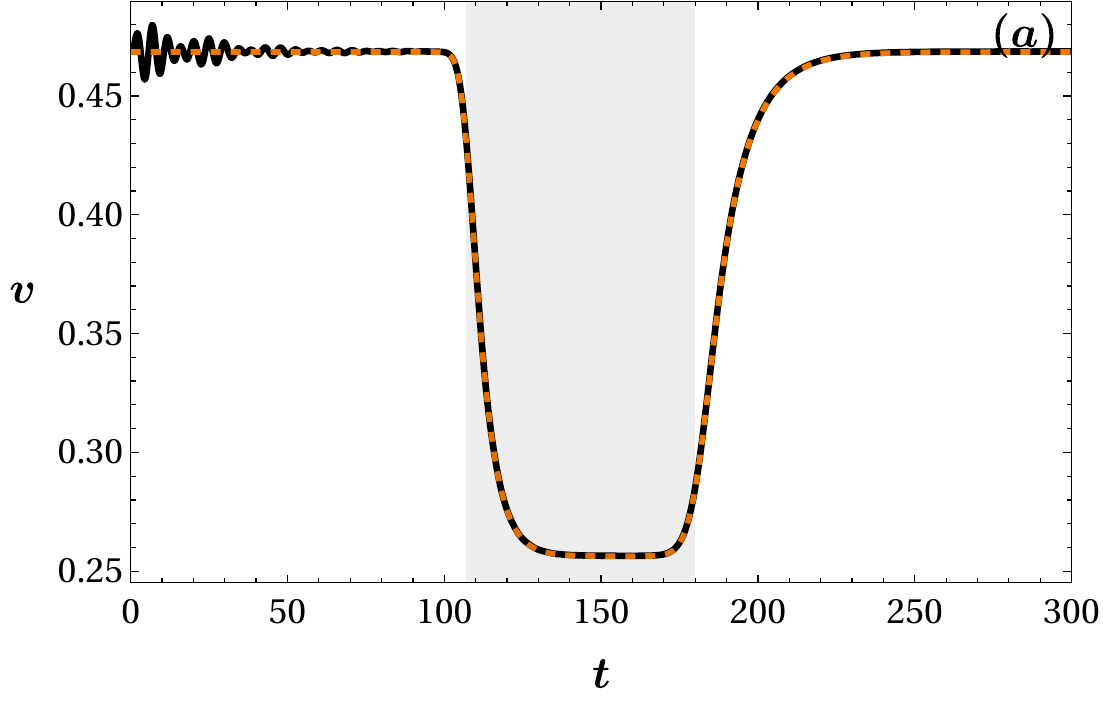}}}
    \quad
    \subfloat{{\includegraphics[height=4.5cm]{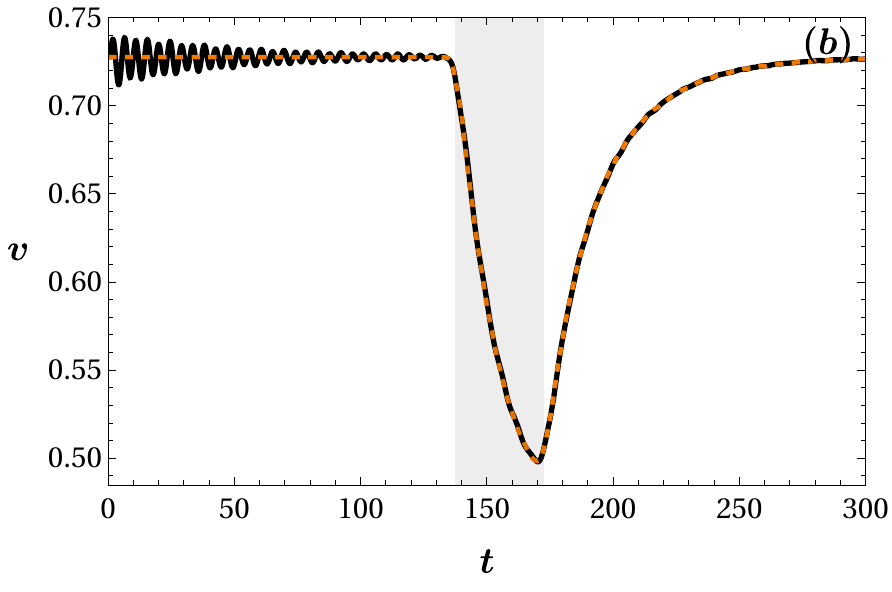}}}
    \caption{Changes in kink velocity when passing through a layer with increased dissipation. 
    In the left panel, the initial position of the kink is $x_0=-5$, while in the right panel it is $x_0=-100$.
    The parameters on the panels are as follows (a) $\eta_0 = 0.1$, $\Gamma = 0.025$,  $\varepsilon = 0.1$ (b) $\eta_0 = 0.04$, $\Gamma = 0.02$,   $\varepsilon = 0.04$. On both panels $\mathtt{L}_0=20$ and $\lambda = 1$. }
    \label{fig_03}
\end{figure}

Figure~\ref{fig_03} shows changes in the velocity of the kink as it passes through a layer with increased dissipation and thickness $\mathtt{L}_0=20$. The parameters on both panels were selected so that the initial speeds of the kink were significantly different, i.e.,  (a) $\eta_0 = 0.1$,  $\Gamma = 0.025$ and (b) $\eta_0 = 0.04$, $\Gamma = 0.02$. As before, speed deviations from the effective model appear at the beginning. Depending on the initial speed, the kink velocity decreases markedly in the layer of increased dissipation. After damping the initial oscillations, the agreement between the effective model and the field model is fairly striking. It can be seen that on both panels, the change in dissipation is at the level of one hundred percent of the initial dissipation.

\begin{figure}[h!]
    \centering
    \subfloat{{\includegraphics[height=4.5cm]{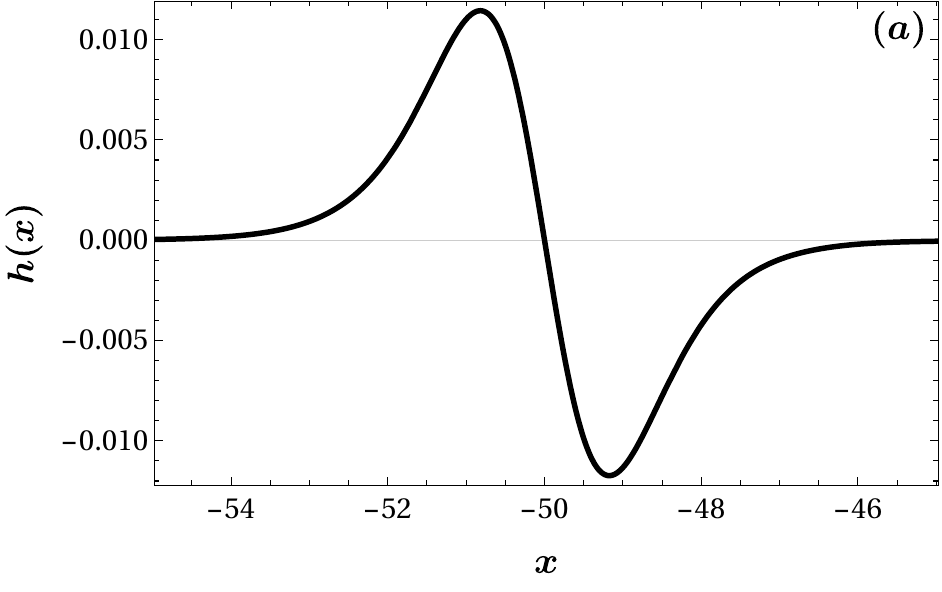}}}
    \quad
    \subfloat{{\includegraphics[height=4.5cm]{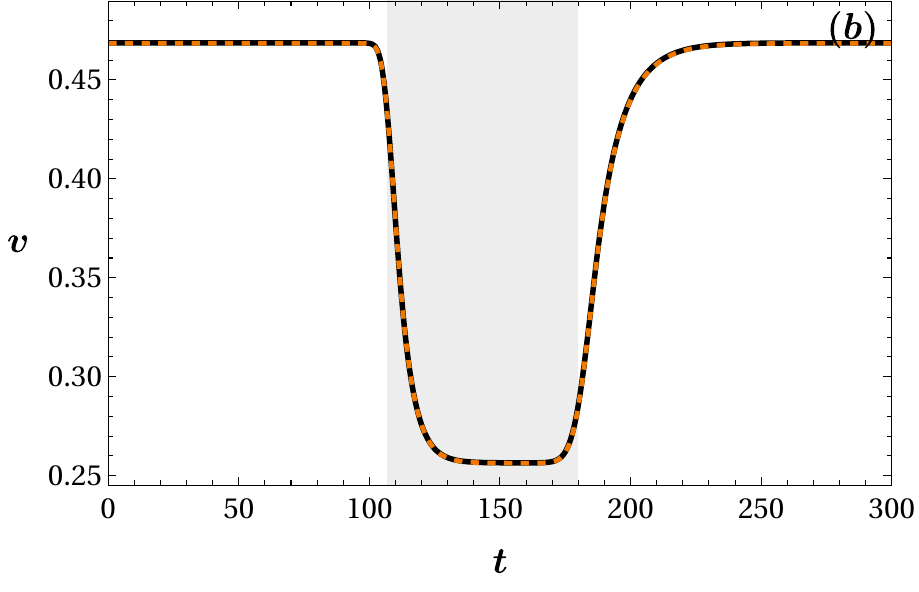}}}
    \caption{ (a) Difference between the gradients of the stationary configuration and the initial condition (after shifting them to the same position) in case of constant dissipation $\eta_0 = 0.1$,   $\varepsilon = 0$ and $\Gamma = 0.025$.  (b) Comparison of the kink velocity from field model (black line) and effective model (orange dashed line). 
    The initial conditions in the field model were chosen based on the stationary solution. The parameters in this panel  are identical to those in Fig.~\ref{fig_03} (a) i.e.,  $\eta_0 = 0.1$, $\varepsilon = 0.1$, $\Gamma = 0.025$  and $x_0=-50$.  }
    \label{fig_04}
\end{figure}

Let us take a closer look at the cause of the oscillations in the initial phase of the motion. We know that, according to the initial conditions \eqref{phi_wp1-kink}, \eqref{phi_wp2-kink}, the initial kink profile is assumed to be in a form that does not account for the presence of forcing (as well as dissipation in the system), while in the system under consideration both effects influence the profile of the kink solution. To verify this hypothesis, we performed the evolution of the initial conditions  in a system with constant values $\Gamma=0.025$, $\eta_0=0.1$ i.e. for $\varepsilon=0$. As in Fig.~\ref{fig_03}, after some time the initial oscillations of the velocity damp out and the motion (due to $\varepsilon=0$) proceeds with a constant velocity. To better illustrate this effect, we computed the gradient of the kink configuration at the final time (here $t_f=200$), corresponding to the stationary kink profile characteristic of the considered system. In the next step, we subtracted the gradient of the initial profile \eqref{phi_wp1-kink} from that of the stationary one
\begin{equation}
    h(x) =  \partial_x \phi(t_f,x+x_c) - \partial_x \phi(0,x).
\end{equation}
 where $x_c=93.7$ describes the shift of the gradient of the final configuration from the final position to the initial position. The resulting difference is shown in Fig. 3 (a). To verify that choosing an appropriate initial condition eliminates the initial oscillations, we insert the configuration obtained at time $t_f$ as the initial condition and obtain the result presented in Fig. 3 (b). It can be seen that in this case no oscillations appear. Moreover, in this case the accuracy of the effective model becomes very high. In Fig.~\ref{fig_04} (b), as before, the solid line represents the velocity obtained from the field model, while the orange dashed line corresponds to the effective model. The parameters in this figure are chosen to be identical to those in Fig.~\ref{fig_03} (a).

Studies concerning the presence of velocity oscillations can be complemented by comparing the spectrum of linear excitations of the kink-type solution with a Fourier analysis of the velocity time series (see  \hyperref[AppendixA]{Appendix A}). The analysis reveals that the oscillations arise from a superposition of a discrete mode and excitations belonging to the continuous spectrum. The presence of a beating pattern in the velocity time course suggests interference between two distinct frequency components, rather than a single oscillatory mode. A possible interpretation is that the discrete mode is nonlinearly coupled to the continuum and excites a quasinormal mode within it, leading to a superposition of both contributions. The gradual decay of the beating envelope can be attributed to both dissipation and energy leakage into the continuum. 

\FloatBarrier

\section{Standard effective descriptions of half-kink}
The second front we address in this paper concerns the half-kink. In this case, external forcing is not necessary to maintain the half-kink motion. Therefore, for simplicity, we assume $\Gamma=0$ in the remaining parts of this paper.
This viewpoint is particularly natural in dissipative systems describing relaxation toward equilibrium, where similar front-like structures arise in phase ordering \cite{Bray1994}, domain growth \cite{Ohta1982}, and reaction–diffusion dynamics \cite{Saarloos2003}.

\subsection{Reformulation of the Lagrangian in the absence of dissipation}
Although the half-kink arises naturally in a dissipative environment (see \hyperref[AppendixB]{Appendix B}), for the purpose of constructing effective models we first consider a version of the system in which dissipation is absent, operating
again at the level of the equation of motion~(\ref{phi4+k}) and
of the Lagrangian~(\ref{L-k}).
To enable the analysis of the half-kink configuration, we introduce a new field variable $\xi = \xi(t, x)$ defined by the equation
\begin{equation}\label{phi_f}
\phi(t,x)= \frac{1}{1+ e^{\xi(t,x)}},
\end{equation}
i.e., once again leveraging an Ansatz featuring the desired asymptotics.
The Lagrangian in the new variables is as follows
\begin{equation}\label{L-xi}
L
=\int_{-\ell}^{+\ell} dx ~ (\partial_{\xi} \phi)^2 \left[ \frac{1}{2}~ (\partial_t
\xi)^2 - \frac{1}{2} ~ (\partial_x \xi)^2 - \frac{1}{4} \lambda \left(2 + e^{\xi} \right)^2
\right].
\end{equation}
This Lagrangian is used in subsequent sections to construct effective models based on a finite number of dynamical degrees of freedom.
\subsection{Model based on classical approach - one degree of freedom }
\begin{figure}[h!]
    \centering
    \subfloat{{\includegraphics[height=4.5cm]{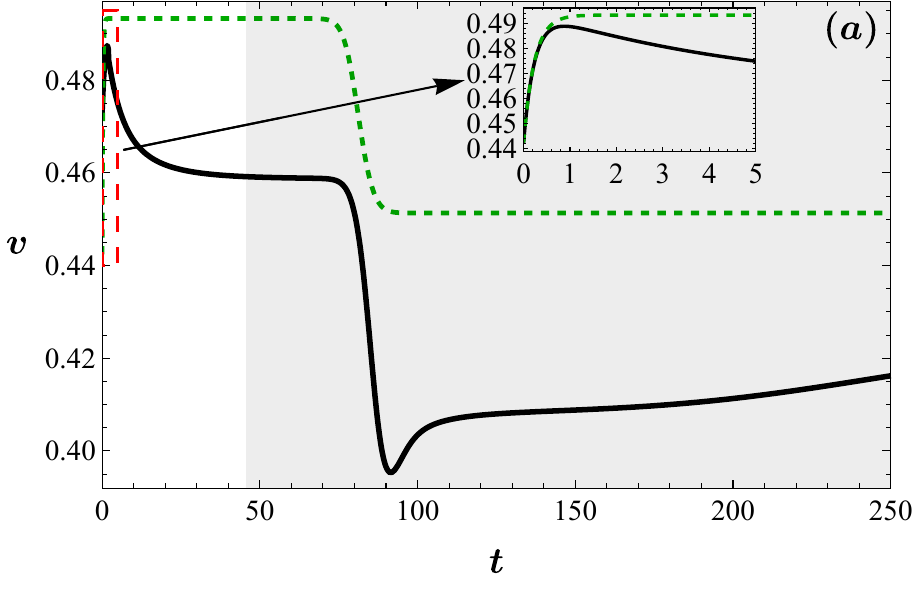}}}
    \quad
    \subfloat{{\includegraphics[height=4.5cm]{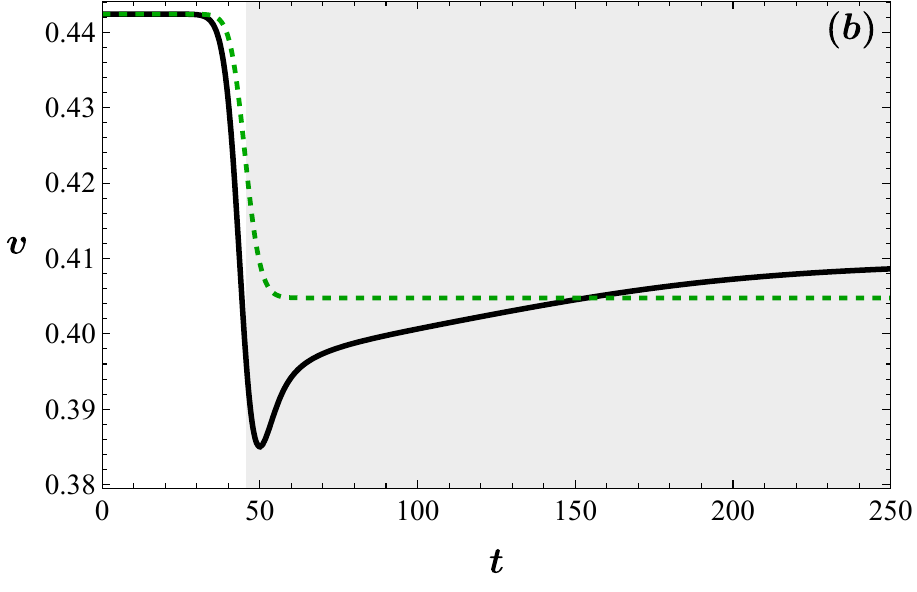}}}
    \caption{Dependence of the velocity on time for a half-kink passing from a region of lower to a region of higher dissipation. The parameters in the figure are $\eta_0=4.3$ and $\varepsilon=0.4$. In panel (a) $\gamma_0=1$ is assumed, while in panel (b) $\gamma_0=\frac{1}{\sqrt{1-v_0^2}}$, where $v_0 = v$ is given by formula \eqref{xi-sol}.}
    \label{fig_05}
\end{figure}

We start constructing effective models, beginning with a single-degree-of-freedom model that describes the position $x_0(t)$ of the half-kink
\begin{equation}
\label{ansatz0}
    \xi(t,x)= \sqrt{\frac{\lambda}{2}} \gamma_0 \left(x - x_0 \right),
\end{equation}
here, $\gamma_0$ corresponds to the characteristic velocity of the half-kink in a medium with constant damping, i.e., determined by the values $\eta_0$ and $\lambda$  in the area where the half-kink is located at the initial moment (see \hyperref[AppendixB]{Appendix B}).  The value of $\gamma_0$ is not a dynamic variable here, but only a constant that allows the behavior of the half-kink to be consistent with the previously chosen initial conditions (we justify this choice in more detail later on). The effective Lagrangian in this case has the form
\begin{equation}\label{Leff-n}
L_{eff} = \frac{1}{2}\, M \Dot{x}_0^2  - V_{eff}.
\end{equation}
The mass coefficient in this formula is determined by the integral $J_0$ given in the \hyperref[AppendixC]{Appendix C}
\begin{equation}
    \label{M_c}
    M = \frac{1}{32} \gamma_0^2 
    ~\lambda \int_{-\ell}^{+\ell} dx \, \sech^4 \frac{\xi}{2} = \frac{1}{16 \sqrt{2}} \, \gamma_0 \sqrt{\lambda}~J_0 .
\end{equation}
In particular, 
in the large-system limit, the mass takes a constant value equal to $M=\frac{1}{6 \sqrt{2}} \sqrt{\lambda} \gamma_0$. While the mass coefficient remains almost constant above a certain size, the effective potential is essentially determined by the value of $\ell$ 
\begin{equation}
    \label{V_c}
   V_{eff} = \frac{1}{64} \lambda \int_{-\ell}^{+\ell} dx \, \sech^4 \frac{\xi}{2} \, \left[ \gamma_0^2 +  \left( 2 + e^{\xi}\right)^2 \right] .
\end{equation}
Integration can be performed explicitly in this case, leading to the expression
\begin{equation}  
\begin{gathered}
    \label{U}
    V_{eff} = \frac{1}{32} \sqrt{\frac{\lambda}{2}}\, J_0 \, \gamma_0 +    
    \frac{\sqrt{\lambda}}{12 \sqrt{2}\, \gamma_0} 
\left[
\frac{1 + 9 e^{\xi_-} + 6 e^{2\xi_-}}{(1 + e^{\xi_-})^3}
- \frac{1 + 9 e^{\xi_+} +6 e^{2\xi_+} }{(1 + e^{\xi_+})^3}
+ 6 \ln \left(\frac{1 + e^{\xi_+}}{1 + e^{\xi_-}} \right)
\right] ,
\end{gathered}
\end{equation}
where $ \xi_+= \sqrt{\frac{\lambda}{2}} \gamma_0 \left(\ell - x_0 \right)$ and $ \xi_-= \sqrt{\frac{\lambda}{2}} \gamma_0 \left(-\ell - x_0 \right).$
In this case, we only have one equation of motion
\begin{equation}
    \label{eff-eq0}
    \frac{d}{d t }\left(\frac{\partial L_{eff}}{\partial \Dot{x}_0} \right) - \frac{\partial L_{eff}}{\partial {x}_0} = \left[ \frac{\partial R_{eff}}{\partial {x}_{-}} -
    \frac{d}{d t }\left(\frac{\partial R_{eff}}{\partial \Dot{x}_{-}} \right) \right]_{PL} .
\end{equation}
 The explicit form of the effective  equation of motion for a system with damping is given by
\begin{equation}
    \label{eq0}
    M \Ddot{x}_0 + \frac{1}{2} \, (\partial_{x_0} M) \, \Dot{x}_0^2 + \partial_{x_0} V_{eff} = - A \Dot{x}_0.
\end{equation}
At the level of the effective description, the function describing dissipation  is defined by means of an integral
\begin{equation}
    \label{B}
    A = \frac{1}{32} \gamma_0^2 ~\lambda \, \int_{-\ell}^{+\ell} dx \, \eta(x) \sech^4 \frac{\xi}{2}.
\end{equation} 
Once again, we see how the integrated overlap of the structure's mass density
with the dissipative profile becomes responsible for the effective dissipation
coefficient of the reduced model.
We tested this highly simplified approach by describing the process of a half-kink front moving from a region with lower dissipation to one with higher dissipation \eqref{eta_function0}. The results of this process are shown in Figure~\ref{fig_05}. Both panels of the figure justify the need to include the factor $\gamma_0$ in the ansatz \eqref{ansatz0}. They depict the changes in the half-kink’s velocity over time as it moves through the system. Note that in panel (a), $\gamma_0$ is absent (i.e. $\gamma_0=1$), which leads to fundamental differences between the velocity evolution obtained from the field model (black solid line) \eqref{phi4+} and the effective model (green dashed line) \eqref{eq0}.
To better illustrate this effect, the part of the graph marked with a red dashed frame has been enlarged within the inset shown in this figure.
On the other hand, the initial conditions in both the field and effective models are entirely consistent in the case shown in panel (b). In the second panel, $\gamma_0$ was adopted for  velocity $v_0=0.44$, which is consistent with the model parameters (see \hyperref[AppendixB]{Appendix B}, 
equation \eqref{xi-sol}). In any case, even in panel (b), the consistency between the effective model and the field model is rather limited.

\subsection{Model based on classical approach - two degrees of freedom }
\begin{figure}[h!]
    \centering
    \subfloat{{\includegraphics[height=4.5cm]{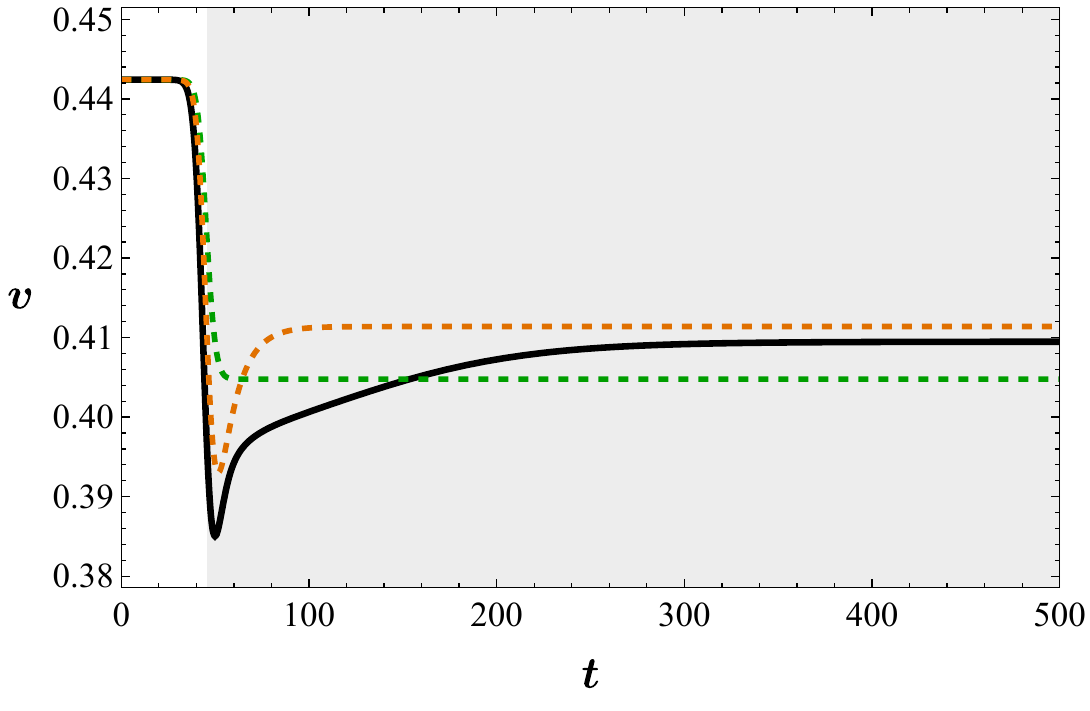}}}
    \caption{Half-kink velocity as a function of time. The black line corresponds to the field model, the orange line to the effective model with two degrees of freedom, and the green line to the reduced model with one degree of freedom. The parameters are chosen as follows: $\eta_0 = 4.3$, $\varepsilon = 0.4$.}
    \label{fig_06}
\end{figure}

In developing an effective two-degree-of-freedom model, we restrict the analytical form of the field $\xi(t,x)$ as follows:
\begin{equation}
\label{xi-s}
    \xi(t,x) =  \sqrt{\frac{\lambda}{2}} \, \gamma(t)
\, (x- x_0(t)) .
\end{equation}
Note that we are dealing with two dynamic degrees of freedom here. The first describes the position of the half-kink $x_0=x_0(t)$, while the second $\gamma=\gamma(t)$ describes the inverse of its thickness. 
 It is worth emphasizing that the second degree of freedom is responsible for describing both kinematic effects (such as the classical Lorentz factor) and dynamic effects related to changes in thickness resulting from interactions present in the system.
In this case, the effective Lagrangian takes the form
\begin{equation}\label{Leff-s}
L^s_{eff} = \frac{1}{2}\, M_s \Dot{x}_0^2 +\frac{1}{2}\, m_s
\Dot{\gamma}^2  + \kappa_s \Dot{x}_0 \Dot{\gamma}   - V_s.
\end{equation}
The mass coefficients that enter the Lagrangian are defined through integrals listed in \hyperref[AppendixC]{Appendix C}.
\begin{equation}
    \label{Ms}
    M_s = \frac{1}{16} \sqrt{\frac{\lambda}{2}} \gamma \, J_0  \stackrel{\ell \to \infty}{=}
    \frac{1}{6} \sqrt{\frac{\lambda}{2}}\gamma, 
    \end{equation}
    \begin{equation}
    \begin{gathered}
    m_s =  \frac{1}{16 \, \gamma^3}  \sqrt{\frac{2}{\lambda}} \, J_2  \stackrel{\ell \to \infty}{=}
    \frac{1}{2 ~\gamma^3} \sqrt{\frac{2}{\lambda}}\left( \frac{\pi^2}{9} - \frac{2}{3} \right) .
    \end{gathered}
\end{equation}
It is evident that in the limiting case, these coefficients take on a considerably simpler form. Notably, the limiting values are reached well within the system sizes considered in this study. The mixed term is preceded by a coefficient that vanishes as $\ell \rightarrow \infty$
    \begin{equation}
    \kappa_s = - \frac{1}{16 \gamma}  \, J_1  \stackrel{\ell \to \infty}{=}
    0 .
\end{equation}
The only quantity that becomes infinite as $\ell \rightarrow \infty$ is the potential, although in a system of finite size this quantity is also finite
\begin{equation}
    \label{vs}
V_s = \frac{1}{64}~\lambda \int_{-\ell}^{+\ell} dx ~ \sech^4 \frac{\xi}{2} ~\left[   \gamma^2 + \left(2 + e^{\xi} \right)^2\right] . 
\end{equation}
Note that in the vicinity of the defect core, i.e. where the field $\phi$ differs from its extreme values $(\phi=0,\phi=1)$, the first term of the potential contribute to the dynamics of the half-kink. The situation is different for the second term, which does not contribute to the dynamics  since this term produces a constant contribution which is dependent only on the size of the system and does not contain any of the dynamical variables. In this sense, the divergence (resulting from this term) in the potential is trivial and can therefore be regularized by subtracting the relevant term.
As before, in the specific case of the field model of Eq.~\eqref{phi4+} and ansatz  \eqref{xi-s} we are considering, the effective model with dissipation that reduces to a system of ordinary differential equations
\begin{equation}
\begin{gathered}
\label{ODEs}
  M_s\Ddot{x}_0 + \kappa_s \Ddot{\gamma} + \frac{1}{2}(\partial_{x_0} M_s) \,\Dot{x}_0^2 + \left(\partial_{\gamma} \kappa_s - \frac{1}{2} \,  \partial_{x_0} m_s \right) \Dot{\gamma}^2 + (\partial_{\gamma} M_s) \Dot{x}_0 \Dot{\gamma}
    +
    \partial_{x_0}V_s = - A_s \Dot{x}_0 - N_s \Dot{\gamma}  ,
  \\  
    m_s\Ddot{\gamma} + \kappa_s \Ddot{x}_0 + \frac{1}{2}  (\partial_{\gamma} m_s) \Dot{\gamma}^2  + \left( \partial_{x_0} \kappa_s - \frac{1}{2} \partial_{\gamma} M_s \right) \Dot{x}_0^2 + (\partial_{x_0} m_s) \Dot{x}_0 \Dot{\gamma}
+\partial_{\gamma}V_s =  - A_{s\gamma} \Dot{x}_0 - N_{s\gamma} \Dot{\gamma}.
\end{gathered}
\end{equation}
The coefficients describing dissipation in the first of the above equations are as follows
\begin{equation}
\begin{gathered}
  A_s =  \frac{1}{32} \, \lambda \gamma^2 \int_{-\ell}^{+\ell} dx \,\eta(x) \sech^4 \frac{\xi}{2} ,
 \quad
 N_s = - \frac{1}{16} \sqrt{\frac{\lambda}{2}}\, \int_{-\ell}^{+\ell} dx \, \eta(x) \, \xi \sech^4 \frac{\xi}{2} 
 .
\end{gathered}  
\end{equation}
In the special case of constant dissipation ($\eta(x)=\eta_0=const$), we obtain 
\begin{equation}
\begin{gathered}
A_s =  \frac{1}{16  }  \sqrt{\frac{\lambda}{2}}\,\gamma \, \eta_0 \, J_0 \stackrel{\ell \to \infty}{=}  \frac{1}{6 } \sqrt{\frac{\lambda}{2}} \,\gamma \, \eta_0 , \quad
N_s= - \frac{1}{16 \gamma} \, \eta_0 \, J_1 \stackrel{\ell \to \infty}{=} 0 .
\end{gathered}  
\end{equation}
Note that in an infinite system, the second coefficient disappears, while the first coefficient is significantly simplified.
The coefficients defining dissipation in the second equation are determined by integrals
\begin{equation}
\begin{gathered}
 A_{s\gamma} =- \frac{1}{16 } \sqrt{\frac{\lambda}{2}} \, \int_{-\ell}^{+\ell} dx \, \eta(x) \, \xi \sech^4 \frac{\xi}{2} =N_s , \quad
 N_{s\gamma} =  \frac{1}{16 \gamma^2} \, \int_{-\ell}^{+\ell} dx \,\eta(x) \, \xi^2 \, \sech^4 \frac{\xi}{2}  .
\end{gathered}  
\end{equation}
If dissipation in the system is constant ($\eta(x)=\eta_0=const$), then the above formulas are greatly  simplified
\begin{equation}
\begin{gathered}
A_{s\gamma}= -\frac{1}{16 \gamma} \, \eta_0 \, J_1 \stackrel{\ell \to \infty}{=} 0 ,\quad
N_{s\gamma} =  \frac{1}{16 \gamma^3} \sqrt{\frac{2}{\lambda}} \, \eta_0 \, J_2 \stackrel{\ell \to \infty}{=}  \frac{1}{ 2 \gamma^3} \sqrt{\frac{2}{\lambda}}\, \eta_0 \left(\frac{\pi^2}{9} - \frac{2}{3} \right) .
\end{gathered}  
\end{equation}
The results of the one- and two-degree-of-freedom models i.e. \eqref{ansatz0} and \eqref{xi-s} are compared with those obtained from the field model \eqref{phi4+} in Figure~\ref{fig_06}. As before, the black solid line corresponds to the field model, the green dashed line to the one-degree-of-freedom model, and additionally  the orange dashed line corresponds to the two-degree-of-freedom model. The system under consideration is described in one region by a dissipation coefficient $\eta_0 = 4.3$, while in the other region the dissipation is increased by $\varepsilon = 0.4$. The gray shaded region marks when the half-kink enters the region of increased dissipation. The half-kink initially moves in the region of lower dissipation and then enters the denser region with a higher dissipation coefficient. During this process, the half-kink initially slows down sharply, after which its velocity gradually increases, stabilizing at approximately $0.41$. In the one-degree-of-freedom model, the half-kink simply reduces its velocity upon entering the region of increased dissipation to a value that is noticeably lower than the final velocity of the half-kink in the field model. Moreover, this model fails to capture the initial significant drop in velocity upon entering the denser region. In the two-degree-of-freedom model, however, the qualitative behavior of the half-kink’s velocity as a function of time is well reproduced. This model captures the pronounced initial drop in velocity upon entering the denser region, followed by an increase in velocity until reaching an asymptotic value, which is, however, slightly higher than the corresponding value obtained in the field model. Additionally, the initial velocity drop in the denser region, although similar in character, is noticeably smaller than in the field model.

\begin{figure}[h!]
    \centering
    \subfloat{{\includegraphics[height=4.5cm]{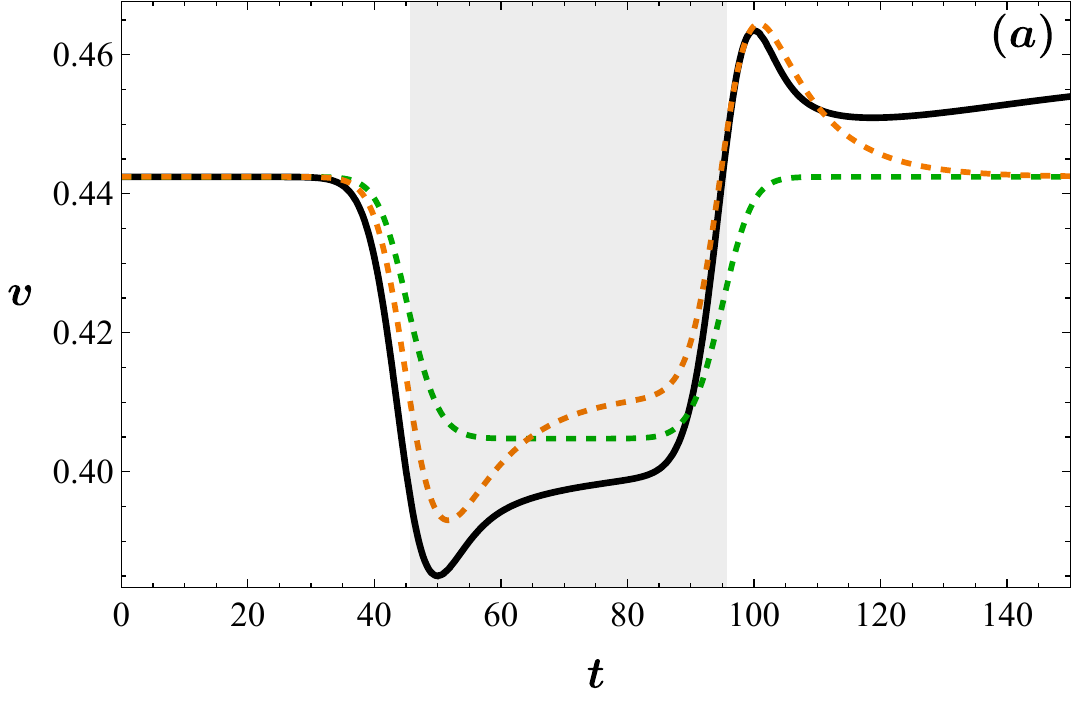}}}
    \quad
    \subfloat{{\includegraphics[height=4.5cm]{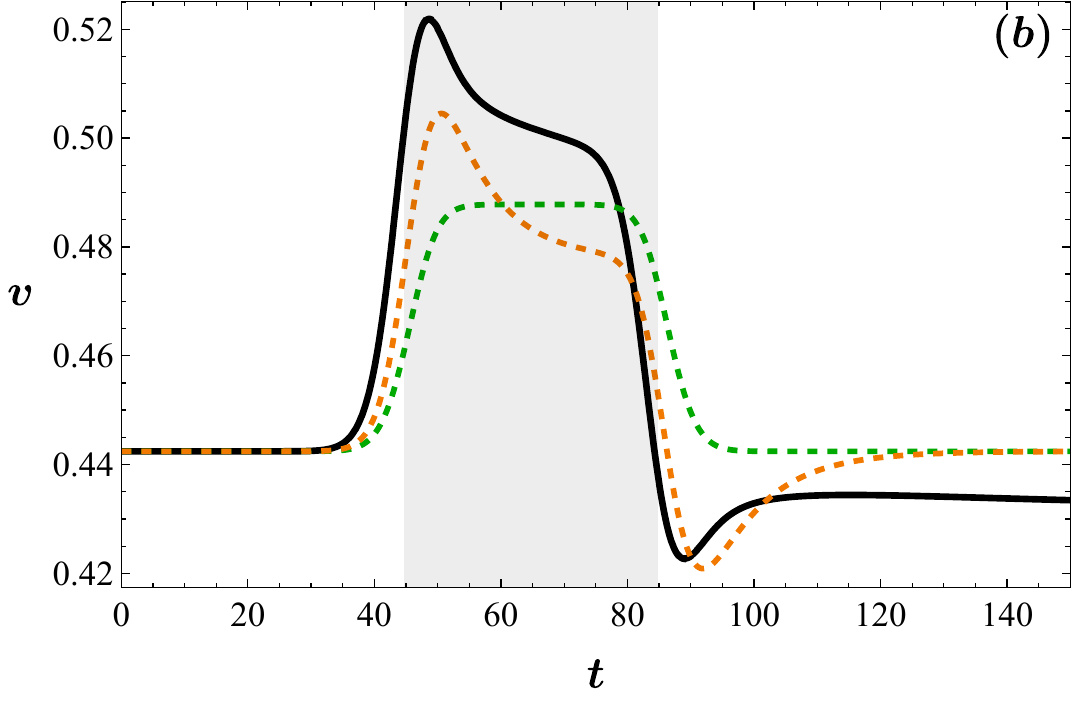}}}
    \caption{Changes in the half-kink velocity during crossing of a layer with modified dissipation. The solid black line represents the result of the field model, the green dashed line the one-degree-of-freedom effective model, and the orange dashed line the two-degrees-of-freedom model.
    Panel (a) corresponds to $\varepsilon = 0.4$, while panel (b) corresponds to $\varepsilon = -0.4$.
    The remaining parameters are equal to $\eta_0 = 4.3$, $\mathtt{L}_0=20$. }
    \label{fig_07}
\end{figure}

The case of a layer with dissipation different from that of the surrounding medium was also investigated. In this case, the function $\eta(x)$ is taken in the form of a compactly supported smooth step-like function \eqref{eta_function}. Figure~\ref{fig_07} shows the velocity of the half-kink as a function of time while it traverses the layer. In panel (a), we observe a temporary decrease in velocity when the dissipation inside the layer is higher than in the surrounding medium. Panel (b) shows the opposite situation, namely an increase in velocity while crossing a layer with lower dissipation than the surrounding environment. In both panels, the solid black line is obtained from the field equation, the green dashed line from the one-degree-of-freedom model, and the orange dashed line from the two-degrees-of-freedom model. The time interval during which the half-kink is located inside the region of modified dissipation is marked as a gray area in the plots. It can be seen that in the one-degree-of-freedom model we observe a simple decrease (panel (a)) or increase (panel (b)) of the velocity within the region of modified dissipation. In contrast, for both the field model and the two-degrees-of-freedom model, the time dependence of the velocity is more complex. Entering the region of modified dissipation produces a rapid jump in velocity, which stabilizes only after some time. This behavior occurs both when entering and when leaving the layer. Such a pattern is visible in both panels (a) and (b). The two-degrees-of-freedom model captures this behavior qualitatively; however, its quantitative agreement with the  field description remains unsatisfactory.

\begin{figure}[h!]
    \centering
    \subfloat{{\includegraphics[height=4.5cm]{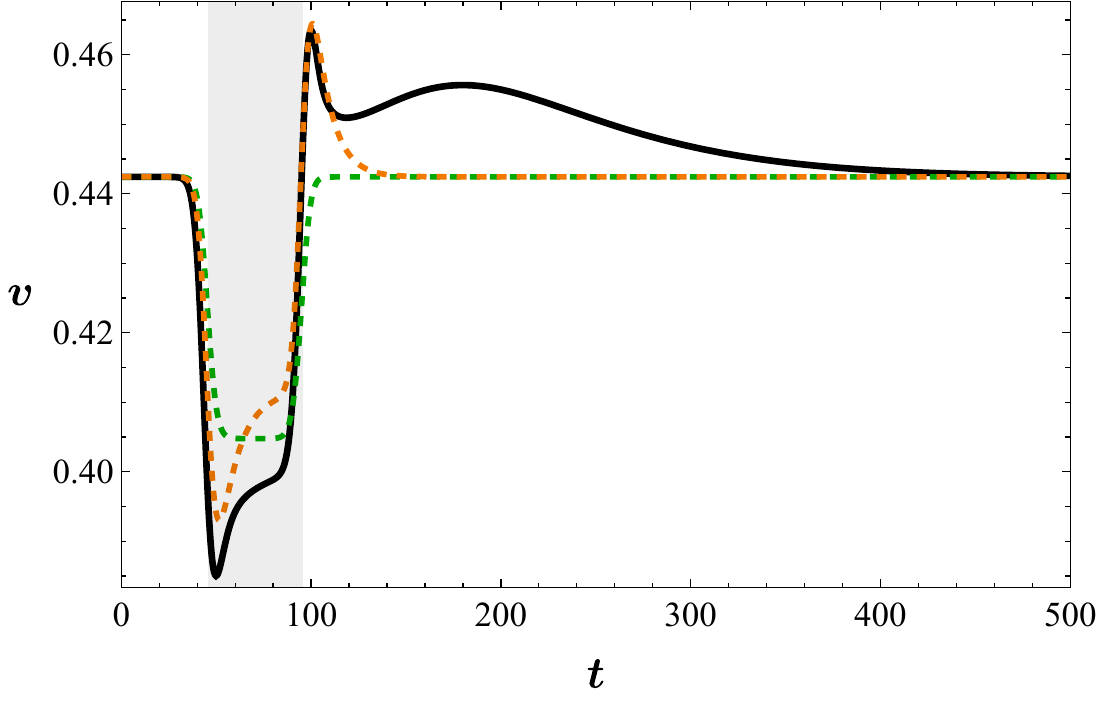}}}
    \caption{Changes in the half-kink velocity over a longer time horizon. The notations are the same as in Figure~\ref{fig_07}. The parameters take the same values as in panel (a) of the previous figure, i.e.
    $\eta_0 = 4.3$, $\mathtt{L}_0=20$ and $\varepsilon = 0.4$.}
    \label{fig_08}
\end{figure}

Figure~\ref{fig_08} shows a process analogous to that presented in Figure~\ref{fig_07}, but over a longer time window to better highlight the relaxation of the velocity after leaving the layer. All notations are the same as in Figure~\ref{fig_07}. It can be seen that, although both effective models correctly reproduce the asymptotic velocity after exiting the layer, the processes leading to this velocity are entirely absent from them. What is particularly striking here is the extremely long time it takes for the velocity to reach a steady state in the field model (after leaving the layer).

\section{Modified effective model of half-kink}
\label{Section5}
Results obtained on the basis of the previous two models raise  the question of whether there exists any model, constructed in the spirit of the models defined in \cite{Dobrowolski2025,Dobrowolski2025a, Dobrowolski2025b,Caputo2024}, that correctly reproduces the time dependence of the half-kink velocity in media with spatially varying dissipation. To investigate this, we define an ansatz containing an unknown function $g(x_0)$
\begin{equation}
\label{xi-n}
    \xi(t,x)= \sqrt{\frac{\lambda}{2}} ~g(x_0) \left(x - x_0 \right).
\end{equation}
Let us note that this third model, like the first one, contains only a single dynamic degree of freedom. Here, we aim to numerically determine the function $g(x_0)$ so as to best reproduce the results of the field model. Taking into consideration the results of the first model,  we will determine the form of the function $\tilde{g}$, which is related to $g$ through a constant factor $\gamma_0$ according to the formula
$g(x_0) = \Tilde{g}(x_0) \gamma_0$. This choice of function $\tilde{g}$ ensures that at the beginning of the evolution it is equal to unity. Due to the position of the function $g$ in the formula \eqref{xi-n}, we chose not to include an independent dynamic variable $\gamma$, but rather
to seek a $g=g(x_0)$. In this new model, the effective Lagrangian takes the form
\begin{equation}\label{Leff-n+}
L_{eff} = \frac{1}{2}\, M \Dot{x}_0^2  - V_{eff}.
\end{equation}
The mass of the half-kink in this model is defined by the integral
\begin{equation}
    \label{M_c+}
   M = \frac{1}{16}\int_{-\ell}^{+\ell} dx \, \sech^4 \frac{\xi}{2} \,\, G_0^2 , \,\,\,\,\,\, G_0 = \frac{\partial_{x_0}g}{g} \, \xi -\sqrt{\frac{\lambda}{2}} g .
\end{equation}
The expression for the mass can be written using the functions defined in \hyperref[AppendixC]{Appendix C}
\begin{equation}
    \label{M1-n}
    M = \frac{1}{16} \sqrt{\frac{2}{\lambda}}\, J_2 \, \frac{1}{g} \left( \frac{\partial_{x_0}g}{g} \right)^2 - \frac{1}{8 } \, J_1 \, \frac{\partial_{x_0} g}{g} + \frac{1}{16} \sqrt{\frac{\lambda}{2}} \, J_0 \,g .
\end{equation}
The effective potential, in turn, takes the form
\begin{equation}
    \label{V_c+}
   V_{eff} = \frac{1}{64} \lambda \int_{-\ell}^{+\ell} dx \, \sech^4 \frac{\xi}{2} \, \left[ g^2 + \left( 2 + e^{\xi}\right)^2 \right].
\end{equation}
Explicit integration over the spatial variable leads to an expression for the effective potential
\begin{equation}  
\begin{gathered}
    \label{U-n}
    V_{eff} = \frac{1}{32 } \sqrt{\frac{\lambda}{2}} \, J_0 \, g(x_0) +    
    \frac{\sqrt{\lambda}}{12 \sqrt{2}\, g(x_0)} 
\left[
\frac{1 + 9 e^{\xi_-} + 6 e^{2\xi_-}}{(1 + e^{\xi_-})^3}
- \frac{1 + 9 e^{\xi_+} +6 e^{2\xi_+} }{(1 + e^{\xi_+})^3}
+ 6 \ln \left(\frac{1 + e^{\xi_+}}{1 + e^{\xi_-}} \right)
\right] ,
\end{gathered}
\end{equation}
where $ \xi_+= \sqrt{\frac{\lambda}{2}}\, g(x_0) \left(\ell - x_0 \right)$ and $ \xi_-= \sqrt{\frac{\lambda}{2}}\, g(x_0) \left(-\ell - x_0 \right).$
The equation describing the dynamics of the single degree of freedom $x_0$ in the presence of dissipation is obtained by using the nonconservative Lagrangian for this system in the framework of the formalism reviewed in the previous sections of the paper
\begin{equation}
    \label{eq1-n}
    M \Ddot{x}_0 + \frac{1}{2} \, (\partial_{x_0} M) \, \Dot{x}_0^2 + \partial_{x_0} V_{eff} = - A \Dot{x}_0 .
\end{equation}
In this equation, the effective dissipation coefficient is defined by the formula
\begin{equation}
    \label{B-n}
    A = \frac{1}{16} \, \int_{-\ell}^{+\ell} dx \, \eta(x) \sech^4 \frac{\xi}{2} \, G_0^2 \, , \,\,\,\,\,\, G_0 = \frac{\partial_{x_0}g}{g} \, \xi -\sqrt{\frac{\lambda}{2}} g.
\end{equation}  
Using this model, we fitted the function $g(x_0)$ to the results of the field model in both cases, namely for the situation where the half-kink moves from a medium with lower damping to one with higher damping, and for the case of crossing a layer with higher damping. It is surprising that, with the adopted ansatz, we are able to (numerically) find the form of the function $g$ that exactly reproduces the relationships obtained within the field model. Furthermore, we are able to propose an explicit analytical form of the function $\tilde{g}$, which perfectly or almost perfectly reproduces the dependence of the speed of the half-kink on time over a wide range of parameters $\eta_0$ and $\varepsilon$. 
 {To obtain the analytical representation $\tilde{g}$, we first evaluated function $g$ on a dense grid in the parameter space, covering $\varepsilon \in (0,1]$ with step $0.02$ and $\eta_0 \in [4,10]$ with step $0.05$. For each pair $(\varepsilon,\eta_0)$, the corresponding parametric form (Eq. \eqref{tilde-g1} or \eqref{tilde-g-2} from \hyperref[AppendixD]{Appendix D} below, depending on the configuration) was fitted, yielding a set of coefficients $a_k(\eta_0,\varepsilon)$. These coefficient dependencies were then approximated by low-degree polynomial functions chosen to achieve an accuracy threshold. The resulting analytical form was subsequently validated over the same parameter ranges of $\varepsilon$ and $\eta_0$ within the framework of the model \eqref{eq1-n} to ensure the accuracy of the approximation.}

\begin{figure}[h!]
    \centering
    \subfloat{{\includegraphics[width=\linewidth]{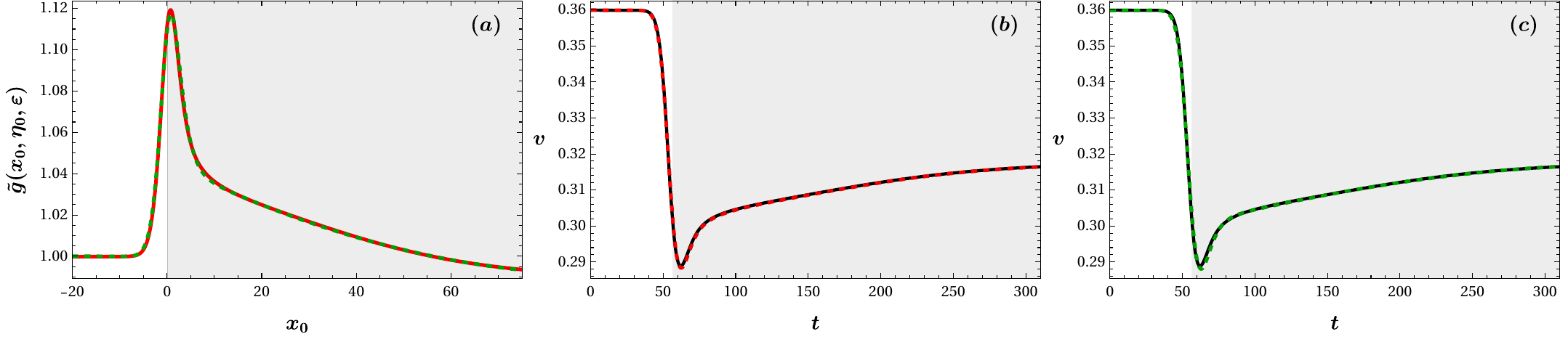}}}
    \caption{ {
    Penetration of a half-kink from a medium with dissipation $\eta_0 = 5.5$ into a medium with dissipation $\eta_0 + \varepsilon$, with $\varepsilon = 0.8$. The figure presents how $\tilde{g}$ varies with $x_0$, along with the time evolution of the half-kink velocity. Panel (a): comparison of numerical and fitted $\tilde{g}(x_0)$ (see \hyperref[AppendixD]{Appendix D}); (b): velocity obtained from the field equation \eqref{phi4+} versus the velocity predicted by the effective model \eqref{eq1-n}, using numerical $\tilde{g}(x_0)$; (c): field-model velocity versus the effective-model result \eqref{eq1-n}, based on the analytical form of $\tilde{g}(x_0)$ (see \hyperref[AppendixD]{Appendix D}). }}
    \label{fig_09}
\end{figure}

Figure~\ref{fig_09} illustrates the process of a half-kink penetrating into a medium with higher dissipation. In this figure, the dissipation in the first medium is taken to be $\eta_0 = 5.5$, and the change in dissipation is fixed at $\varepsilon = 0.8$, i.e.,
the dissipation varies from $\eta_0$ to $\eta_0+\varepsilon$. The gray area represents the time during which the half-kink is in a region of modified dissipation. Panel (a) compares the function $\tilde{g}$ obtained numerically (red line) with an analytical fit to this numerical profile (green dashed line). The function $\widetilde{g}$ is determined numerically so as to best reproduce $x_0$ obtained as a solution of Eq.~\eqref{eq1-n}. In turn, its analytical form is fitted (within a “basis” composed of the half-kink profile and its derivatives evaluated at $x_0$ and $x_0 - \ell$) so as to best match the numerical representation of $\widetilde{g}$. As shown in panel (a), the analytical fit presented in \hyperref[AppendixD]{Appendix D} reproduces the numerical profile of $\widetilde{g}$ remarkably well.  In panel (b), we also present a comparison of the velocity–time dependence obtained from the field model \eqref{phi4+} (solid black line) and from the effective model \eqref{eq1-n}, where the numerical form of the function $\widetilde{g}$ has been implemented (red dashed line). 
No difference between the two curves can be detected with ``naked eye''.
Panel (c), for completeness, presents a comparison between the velocity–time dependence obtained from the field model \eqref{phi4+} (solid black line) and that obtained from the effective model \eqref{eq1-n}, where the analytical form of the function $\widetilde{g}$ (see \hyperref[AppendixD]{Appendix D}) has been implemented (green dashed line). The agreement here is also very good. It is worth emphasizing that, although the figure shows these dependencies for a particular choice of $\eta_0$ and $\varepsilon$, the analytical form presented in \hyperref[AppendixD]{Appendix D} reproduces the velocity–time dependence equally well across a broader range of parameters. Our analysis shows that for $\varepsilon \in (0,1]$ and $\eta_0 \in [4,10]$, the analytical model defined by Eqs.~\eqref{phi_f} and \eqref{xi-n}, together with the form given in \hyperref[AppendixD]{Appendix D}, reproduces the velocity–time dependence of the half-kink very well when used in Eq.~\eqref{eq1-n}.
It is worth emphasizing that, in the present study of the half-kink dynamics, the field model corresponds to Eq.~\eqref{phi4+} with $\Gamma = 0$. As illustrated in the figures, once the half-kink enters the more dissipative medium, its velocity drops sharply and then gradually recovers, approaching an asymptotic value.

The next figure, namely Figure ~\ref{fig_10}, presents the results obtained for a layer with increased dissipation. In the figure, the thickness of the layer with modified dissipation is taken to be $\mathtt{L}_0 = 40$. Within the layer itself, the dissipation coefficient increases by $\varepsilon=1$ (again to $\eta_0 + \varepsilon$), whereas outside the layer the dissipation is $\eta_0=6$. The shaded region indicates the time during which the half-kink is within the layer of modified dissipation. Panel (a) shows a comparison of the function $\widetilde{g}$ (solid red line), obtained numerically so that the effective model \eqref{eq1-n} reproduces the velocity–time dependence derived from the field model \eqref{phi4+}, and the analytical fit of this function presented in \hyperref[AppendixD]{Appendix D} (green dashed line).
Panel (b) presents a comparison of the velocity–time dependence of the half-kink obtained from the field model \eqref{phi4+} (solid black line) and that obtained from equation \eqref{eq1-n} based on the numerical form of the function $\widetilde{g}$ (red dashed line). Naturally, this fit is very good for arbitrarily chosen parameter values.
Panel (c), in turn, shows a comparison of the velocity obtained from the field model (solid black line) with the velocity obtained in the effective model based on the analytical form of the function $\widetilde{g}$ according to formula in \hyperref[AppendixD]{Appendix D} (green dashed line).  It is evident that the numerical fits in the figures are essentially perfect, while the analytical ones are very good, both in terms of $\tilde{g}(x_0)$ and the time dependence of the velocity. It is worth emphasizing that the analytical form presented in \hyperref[AppendixD]{Appendix D} reproduces the results of the field model very well over a fairly wide range of parameter values, i.e., $\varepsilon \in (0,1]$ and $\eta_0 \in [4,10]$.
The course of the process of a half-kink passing through a layer with increased dissipation, shown in panels b) and c), demonstrates that upon entering the layer, a rapid deceleration occurs. Subsequently, within the layer, the velocity approaches a stationary value corresponding to the dissipation $\eta_0+\varepsilon$. After leaving the layer, the velocity increases quite rapidly. Asymptotically (for large times), the stationary velocity characteristic for the dissipation coefficient $\eta_0$ is reached.
In the case of passing through the layer (after exiting it), a somewhat intriguing increase in velocity is observed. For the parameters used in this figure, it occurs for times $t \in (240-300)$. It is worth noting that this phenomenon was always observed (for intervals depending on the choice of parameters) whenever the half-kink exited the layer with modified dissipation. Presently, we do not have a 
theoretical explanation for this type of fluctuation. Developing
an intuitive understanding thereof is an interesting direction for
further study.

\begin{figure}[h!]
    \centering
    \subfloat{{\includegraphics[width=\linewidth]{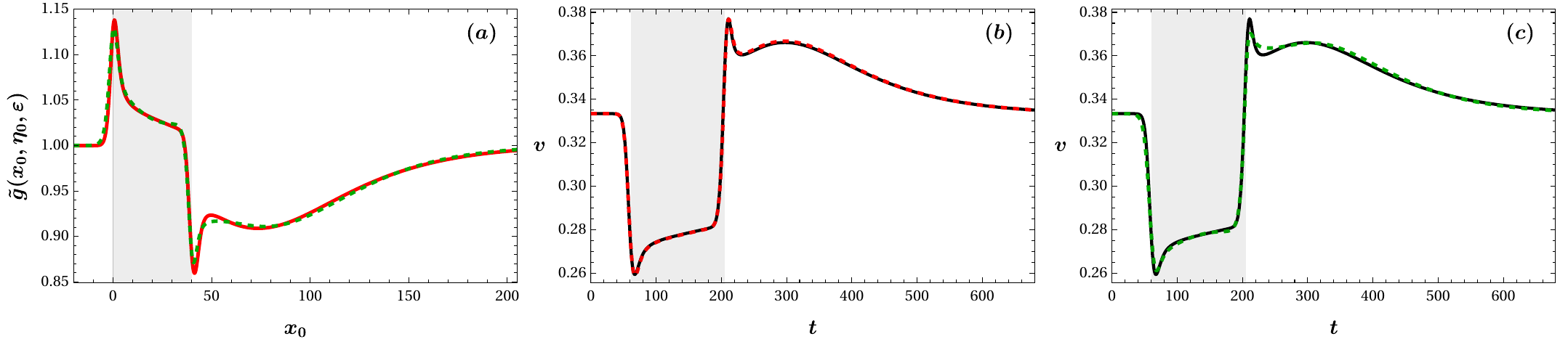}}}
    \caption{The passage of a half-kink through a layer of increased dissipation for $\mathtt{L}_0 = 40$, $\eta_0 = 6$ and $\varepsilon = 1$. Panel (a): the function $\tilde{g}(x_0)$ -- numerical result (solid red line) and analytical fit (green dashed line). Panel (b): the half-kink velocity obtained from the field model (solid black line) and the velocity obtained using the numerical form of the function $\tilde{g}$ (red dashed line). Panel (c): the half-kink velocity obtained from the field model (solid black line) and  obtained using the analytical approximation of the function $\tilde{g}$ (green dashed line).}
    \label{fig_10}
\end{figure}

\FloatBarrier

\section{Conclusions \& Outlook}
In this work, we investigated the influence of dissipation present in the system on the propagation of front-like solutions. In the first part, we focused on the motion of configurations in the form of a kink. Based on the framework for modeling dissipative systems developed in the articles \cite{Galley2013,Kevrekidis2014}, we constructed a model with two dynamical degrees of freedom. The Ansatz underlying our construction is entirely standard and incorporates two dynamical variables: the position of the kink and the inverse of its width.
We considered two scenarios of kink motion. In the first, the kink enters a region of enhanced dissipation. The presence of increased dissipation naturally leads to a reduction in the kink’s velocity. In our setup, sustained motion of the kink was made possible by the presence of an external driving force. In fact, the motion always occurred with a velocity determined by the balance between dissipation and driving. The model we proposed provided a very accurate description of all velocity changes.
Similarly good agreement was obtained for the case in which the kink passes through a layer with modified dissipation. In both cases, however, deviations from the model appeared at the initial stage of the motion and manifested themselves as damped oscillations of the velocity. The origin of these oscillations was identified. Their source lies in the fact that the function used as the initial condition, corresponding to the model with $\Gamma = 0$ and $\eta(x) = 0$, is not an exact solution when nonzero driving and dissipation are present in the system.
This discrepancy between configurations was demonstrated as follows. Within the field model, for fixed values of dissipation and driving, we evolved the aforementioned initial condition. After some time (once the oscillations had decayed), we obtained the true numerical solution, whose velocity no longer exhibited oscillations. By shifting this solution back to the initial position, we were able to compare it with the original initial condition and thereby reveal a small but definite difference.
Moreover, by analyzing the excitation spectrum of the obtained numerical solution, we were able to compare it with the Fourier analysis of the velocity oscillations observed at the beginning of the evolution. This comparison shows that the observed oscillations correspond to the discrete mode as well as to a value belonging to the continuous spectrum of excitations of the kink solution.
Finally, by using the numerical solution as the initial condition, we demonstrated that the oscillations are completely eliminated.

We carried out analogous considerations for a front in the form of a half-kink, i.e., a solution that interpolates between the unstable configuration $\phi = 0$ and the stable configuration $\phi = 1$. As before, we considered identical dissipative settings, namely two media characterized by different dissipation coefficients and a layer with modified dissipation.
In the case of the half-kink, an important limitation arises from the requirement of finite energy, which led us to restrict the system to a finite spatial domain.
While this technicality ensures that no divergence
arises in the model, the nature of the phenomenology suggests
that a suitable renormalization of the energy will yield 
identical results even for the ---in principle--- infinite
domain case.
We employed the same methodology for constructing effective models as in the kink case, namely one based on a nonconservative Lagrangian. In this way, we obtained two models: one with a single degree of freedom and another with two degrees of freedom.
It turns out that, in this case, both effective models yield results that deviate significantly from the predictions of the  field model. The discrepancy is particularly pronounced for the model with a single degree of freedom. The two-degree-of-freedom model reproduces, to some extent, at least qualitatively, the time dependence of the velocity; however, the quantitative differences remain substantial. Moreover, in the case of a layer with modified dissipation, even the two-degree-of-freedom model fails 
significantly to reproduce the process by which the system approaches a stationary velocity in the field model after exiting the layer.
This naturally raises the question of whether an effective model exists that can faithfully reproduce the results of the field description. In the final part of this work, we demonstrate that such a model can indeed be constructed, even with a single degree of freedom, based on an unknown function $g(x_0)$, which we then proceed
to ``discover'' through suitable optimization methods. Analytical expressions for this function were derived for both the case of two media with different dissipation coefficients and for the layered configuration. The proposed forms of $g(x_0)$ allow one to reproduce the behavior of the half-kink over a fairly broad range of parameters $\eta_0$ and $\varepsilon$ (namely, $\varepsilon \in (0, 1]$ and $\eta_0 \in [4, 10]$).

The present work demonstrates that, while the modeling of kink dynamics in media with spatially varying dissipation is relatively well understood (although, there too,
some suitable adjustments to the collective coordinate approach need to be implemented), the situation is far less clear in the case of the half-kink. In particular, a number of open questions remain concerning both the construction of effective models and their
interpretability.
A key result in this context comes from numerical simulations performed within the field framework, which revealed the existence of an unexpected effect: an extremely long relaxation time of the half-kink velocity toward its equilibrium value after passing through a layer of increased dissipation. Understanding the mechanisms responsible for such behavior constitutes a significant research challenge. Furthermore, it does
not escape us that the studies of kinks (and half-kinks) in such damped
(and occasionally driven) settings have been largely limited to one-dimensional
settings in space. A challenge ---starting from radial generalizations and 
extending to genuine higher-dimensional settings--- involves developing an understanding
of the generalization of the relevant phenomenology starting from two-dimensional
settings. Such studies are currently in progress and will be reported in future 
publications.

\FloatBarrier
\section*{Acknowledgments}
We gratefully acknowledge Polish high-performance computing infrastructure PLGrid (HPC Center: ACK Cyfro\-net AGH) for providing computer facilities and support within computational grant no. PLG/2025/018503 (JG).

\clearpage

\appendix
\section{Appendix}
\label{AppendixA}

\setcounter{equation}{0}
\renewcommand{\theequation}{\thesection\arabic{equation}}
\renewcommand{\theHequation}{\thesection\arabic{equation}}

To begin, we will consider the effect of dissipation on the spectrum of linear excitations of the kink in the $\phi^4$ model in the absence of external forcing (thus we set $\Gamma = 0$). Here, we follow the approach presented in \cite{Demirkaya2014}. In this case, the equation of motion takes the form
\begin{equation}
\label{eq-const-eta0}
    \partial_{t}^2 \phi 
+ \eta_0  \, \partial_{t}  \phi
- \partial_{x}^2 \phi
+ \lambda \, \phi (\phi^2 - 1)
= 0 .
\end{equation}
Since the static sector of the model is identical in the presence and absence of dissipation, the analysis concerns the stability of the standard kink solution $\phi_K$, present in both versions
$\phi(t,x)=\phi_K(x) + \delta \phi(t,x),$ where $\phi_K$ satisfies the equation
\begin{equation}
\label{eq-kink}
- \partial_{x}^2 \phi_K
+ \lambda \, \phi_K (\phi_K^2 - 1)
= 0 .
\end{equation}
The analytical form of this solution is well known
$\phi_K(x) = \tanh \sqrt{\frac{\lambda}{2}} (x-x_0).$ In turn, the linear perturbation satisfies the equation
\begin{equation}
\label{eq-delta0}
    \partial_{t}^2 \delta\phi 
+ \eta_0 \partial_{t} \delta \phi  + \widehat{H}\, \delta \phi
= 0 ,
\end{equation}
where $\widehat{H}$ is the operator describing the quantum Hamiltonian with the Pöschl–Teller potential
\begin{equation}
    \label{H}
    \widehat{H} = -  \partial_{x}^2   +\lambda \, \big(3 \, \phi_K^2 - 1\big) .
\end{equation}
The eigenfunctions ($\widehat{H} \psi = E \psi$) of this operator are well known, and its spectrum consists of two discrete eigenvalues, $E_0=0$ and  $E_1=\frac{3 \lambda}{2}$, as well as a continuous part starting at $2 \lambda$ i.e. $E_k \geq 2 \lambda$. Based on this observation, we can carry out a stability analysis of the kink in the presence of dissipation by substituting $\delta \phi = e^{\Lambda t} \psi(x)$ into equation \eqref{eq-delta0}.
As a result, the aforementioned equation reduces to the form
\begin{equation}
\label{eq-deigen0}
    \Lambda^2 \psi 
+ \eta_0 \Lambda \psi  + \widehat{H}\, \psi
= 0 .
\end{equation}
Using the knowledge of the spectrum of the operator $\widehat{H}$, we can obtain an analytical form of the exponents
\begin{equation}
    \label{lambda0}
    \Lambda_{\pm} =  - \frac{\eta_0}{2} \pm \sqrt{\frac{\eta_0^2}{4} -  E} .
\end{equation}
To begin with, the eigenvalue $E_0 = 0$ of the operator $\widehat{H}$ corresponds to the translational mode and its partner describing damping
\begin{equation}
    E_0=0 \quad \Rightarrow  \quad \Lambda_{+} = 0, \qquad \Lambda_{-} = - \eta_0 .
\end{equation}
The second discrete eigenvalue, depending on the damping present in the system, corresponds either to damped oscillations with frequency $\omega$ (for weak dissipation)
\begin{equation}
\label{omega-dis}
    E_1=\frac{3 \lambda}{2}  \quad \mathrm {and} \quad \eta_0 < \sqrt{6 \lambda}\quad \Rightarrow  \quad \Lambda_{\pm} = - \frac{\eta_0}{2} \pm i \omega, \qquad \omega = \sqrt{\frac{3}{2} \,\lambda - \frac{\eta_0^2}{4}} \, ,
\end{equation}
or to pure damping (when the dissipation is large). In the latter case, both exponents are negative
\begin{equation}
    E_1=\frac{3 \lambda}{2}  \quad \mathrm {and} \quad \eta_0  \geq \sqrt{6 \lambda}\quad \Rightarrow  \quad \Lambda_{\pm} = - \frac{\eta_0}{2}  \left( 1 \pm  \sqrt{1 - \frac{6 \lambda}{\eta_0^2}}\right) < 0 . 
\end{equation}
Likewise, in the case of the continuous spectrum, for weak dissipation, we are dealing with damped oscillations with frequencies $\omega_k$
\begin{equation}
    E_k=2 \lambda + k^2  \quad \mathrm {and} \quad \eta_0  < 2 \sqrt{2 \lambda}\quad \Rightarrow  \quad \Lambda_{\pm} = - \frac{\eta_0}{2}  \pm i \omega_k, \qquad \omega_k = \sqrt{2 \lambda + k^2  - \frac{\eta_0^2}{4}} \, .
\end{equation}
If the dissipation is large, then at the edge of the continuous spectrum ($k = 0$) we obtain pure damping. On the other hand, for sufficiently large $k$, oscillations appear
\begin{equation}
    E_k=2 \lambda + k^2  \quad \mathrm {and} \quad \eta_0  \geq 2 \sqrt{2 \lambda}\quad \Rightarrow  \quad \Lambda_{\pm} = - \frac{\eta_0}{2}  \pm \frac{1}{2} \, \sqrt{\eta_0^2 - 8 \lambda - 4 k^2} \, .
\end{equation}
It is worth noting, however, that the real part of $\Lambda_{\pm}$ is always negative. In summary, in the presence of dissipation, all excitations are stable.

Next, we  examine the linear stability of a kink-type solution in the presence of external forcing and dissipation as well. In the case considered, we  assume both a constant value of $\Gamma$ and of the coefficient $\eta_0$. The wave equation in this case takes the form
\begin{equation}
\label{eq-const-eta1}
    \partial_{t}^2 \phi 
+ \eta_0  \, \partial_{t}  \phi
- \partial_{x}^2 \phi
+ \lambda \, \phi (\phi^2 - 1)
= -\Gamma .
\end{equation}
It is worth emphasizing that in this model there exists a kink-type solution traveling at
constant speed whose velocity is determined by the balance between the external forcing and dissipation. The velocity of such a kink, as obtained from the effective model \eqref{2dof_ansatz-k}, is given by
\begin{equation}
\label{v-s-kink}
    v = \frac{1}{\sqrt{1+ \frac{2 \lambda \eta_0^2}{9 \Gamma^2}}} .
\end{equation}
To investigate the stability of such a solution, we perform the transformation $t'=t, \, x'=x-v t$ to the rest frame of the kink.
In co-moving coordinates, the field equation reads as follows
\begin{equation}
\label{eq-comooving}
    \partial_{t'}^2 \phi - 2 v \partial_{t'} \partial_{x'} \phi
- (1-v^2) \partial_{x'}^2 \phi
+ \eta_0 \partial_{t'} \phi - \eta_0 v \partial_{x'} \phi +\lambda \, \phi (\phi^2 - 1)
= -\Gamma .
\end{equation}
The vacuum solutions in this model are solutions of an algebraic equation
\begin{equation}
    \label{vac}
    \lambda \, \phi_{\pm} (\phi_{\pm}^2 - 1)
= -\Gamma .
\end{equation}
It is worth noting that two minima exist in the model as long as the condition $|\Gamma|< \frac{2 \lambda}{3 \sqrt{3}}$ is satisfied, and they take the form (approximation is for small values of $\Gamma/\lambda$)
\begin{equation}
    \phi_{-} = \frac{2}{\sqrt{3}} \cos \left( \frac{2 \pi - \beta}{3}\right)\approx -1 - \frac{\Gamma}{2 \lambda}, \qquad \phi_{+} = \frac{2}{\sqrt{3}} \cos \left( \frac{\beta}{3}\right)\approx + 1 - \frac{\Gamma}{2 \lambda} , \quad \beta = \arccos \left( -\frac{3 \sqrt{3} \, \Gamma}{2 \lambda}\right) \, .
\end{equation}
The corresponding masses are as follows
\begin{equation}
    m_{\pm}^2 = \lambda (3 \phi_{\pm}^2 -1) \approx 2 \lambda \mp 3 \Gamma .
\end{equation}
Moreover, there exists a solution $\phi_0(x')$ that closely resembles the form of the kink. We are, in fact, interested in the stability of this solution.
For this reason, we perform an expansion $\phi(t',x')=\phi_0(x') + \delta \phi(t',x')$ around the static, in new coordinates, solution $\phi_0(x')$.
 The function $\phi_0$ satisfies the equation
\begin{equation}
\label{eq-zero}
- (1-v^2) \partial_{x'}^2 \phi_0
- \eta_0 v \partial_{x'} \phi_0 +\lambda \, \phi_0 (\phi_0^2 - 1)
= -\Gamma .
\end{equation}
This equation differs from the one satisfied by the standard static kink, which is obtained under the assumptions $\Gamma = 0$, $\eta_0 = 0$, and $v = 0$. Consequently, the profile of the solution $\phi_0$ differs from the standard kink form.
At first order in the expansion with respect to $\delta \phi$, we obtain
\begin{equation}
\label{eq-delta}
    \partial_{t'}^2 \delta\phi - 2 v \partial_{t'} \partial_{x'} \delta \phi
- (1-v^2) \partial_{x'}^2 \delta \phi
+ \eta_0 \partial_{t'} \delta \phi - \eta_0 v \partial_{x'} \delta \phi +\lambda \, \big(3 \, \phi_0^2 - 1\big) \, \delta \phi
= 0 .
\end{equation}
By factoring out the time dependence in exponential form $\delta \phi (t',x') = e^{\Lambda t'} \psi(x')$, we obtain 
\begin{equation}
\label{eq-Lambda+}
 - (1-v^2) \partial_{x'}^2  \psi  - (2 v  \Lambda + \eta_0 v ) \partial_{x'}  \psi +   \Lambda^2 \psi
+ \eta_0  \Lambda \psi  +\lambda \, \big(3 \, \phi_0^2 - 1\big) \, \psi
= 0 .
\end{equation}
The lower bound of the continuous spectrum can be determined by analyzing the asymptotic behavior of the kink as $x \rightarrow \pm \infty$, i.e., $\phi_0 \rightarrow \phi_{\pm}$. Considering Fourier modes of the form $\psi(x') = e^{i k x'}$, one obtains an algebraic equation for the exponents $\Lambda$
\begin{equation}
\label{Lambda-equ}
    \Lambda^2 +( \eta_0 - 2 i v k) \Lambda + (1-v^2) k^2 - i \eta_0 v k + m^2_{\pm} =0 .
\end{equation}
The solutions of this equation are as follows
\begin{equation}
    \label{omega-cont}
    \Lambda_{\pm} = - \frac{\eta_0}{2} + i v k \pm \, \sqrt{\frac{\eta_0^2}{4} - ( k^2 + m^2_{\pm})} \, .
\end{equation}
As before, we are dealing with two regimes. In the first, where the dissipation is sufficiently small $\eta_0< 2 m_{\pm}$, we observe oscillatory modes damped by the factor $-\eta_0/2$ (at the boundary $k=0$) 
\begin{equation}
    \Lambda_{\pm} = - \frac{\eta_0}{2} \pm i \omega_c, \qquad \omega_c = \sqrt{m^2_{\star} - \frac{\eta_0^2}{4}} = \sqrt{2 \lambda - 3 \Gamma - \frac{\eta_0^2}{4}}, \qquad m^2_{\star} = \min\{m_+^2,m_-^2\}.
\end{equation}
In the case where the dissipation is sufficiently large $\eta_0 \geq 2 m_{\pm}$, we are dealing with pure damping (as before $k=0$) 
\begin{equation}
    \Lambda_{\pm} = - \frac{\eta_0}{2} \pm \frac{1}{2}\sqrt{\eta_0^2 - 4 m^2_{\pm}} < 0 .
\end{equation}
Thus, we are dealing with stable excitations for $m^2_{\pm}>0$ , i.e., whenever the potential has two minima.

As in the previous model, in the case $\Gamma \neq 0$, a zero mode appears in addition to the continuous spectrum.
We can demonstrate this by differentiating equation \eqref{eq-zero} with respect to $x'$ 
\begin{equation}
    - (1 - v^2) \partial^2_{x'} \phi'  - \eta_0 v \partial_{x'} \phi' + \lambda (3 \phi_0^2 - 1) \phi'=0 .
\end{equation}
In the above equation, the prime denotes differentiation with respect to the variable $x'$. This equation coincides with \eqref{eq-Lambda+} for $\Lambda = 0$ and $\psi = \phi'$. Consequently, the eigenvalue $\Lambda = 0$ is associated with the eigenfunction $\psi = \phi'$.

The final mode we comment on is a damped, discrete oscillatory mode that is present in the regime of weak dissipation. In this case, one may employ perturbation theory with respect to the parameter $\Gamma$, which is assumed to be small throughout this work. The base solution is given by $\phi_K(x')$. In equation \eqref{eq-Lambda+}, we expand $\Lambda = \Lambda^{(0)} + \Gamma \Lambda^{(1)}$ and $\psi = \psi^{(0)} + \Gamma \psi^{(1)}$. It follows that the first-order correction to $\Lambda$ vanishes, while the second-order contribution in $\Gamma$ remains negligibly small within the present analysis. For this reason, to linear order in $\Gamma$, we can use Eq. \eqref{omega-dis} to determine the frequency of this mode.

\begin{figure}[h!]
    \centering
    \subfloat{{\includegraphics[height=4.75cm]{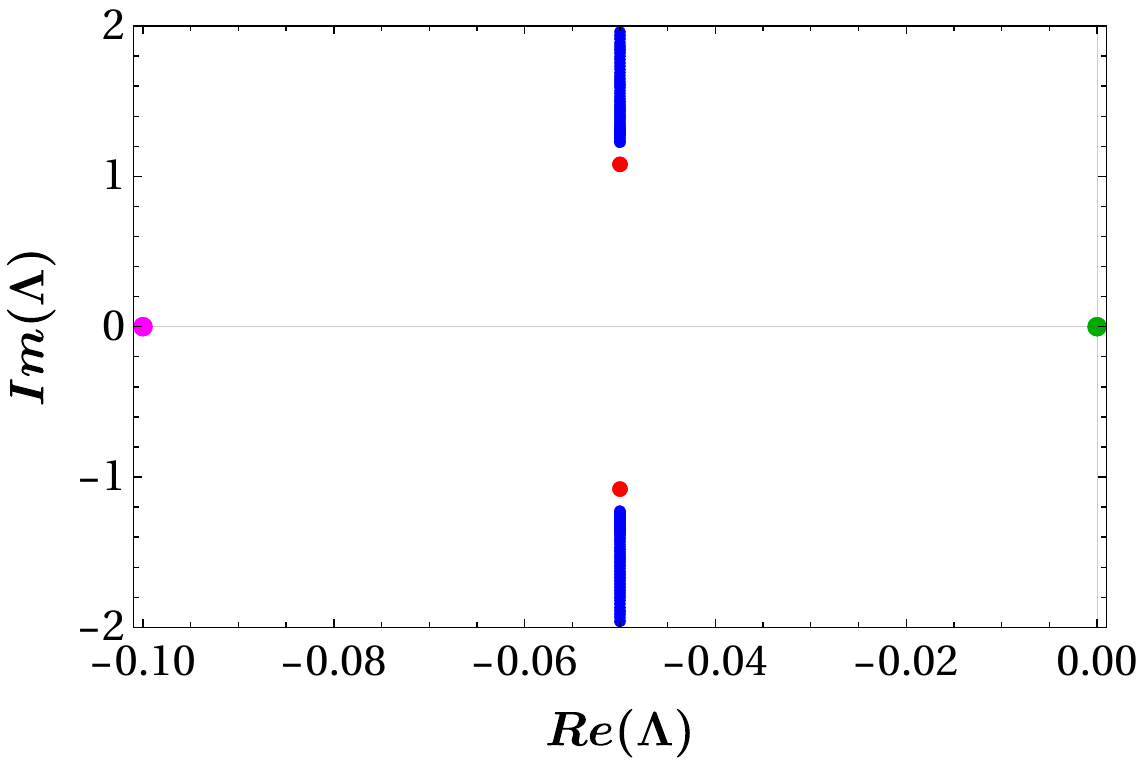}}}
    \caption{The complex plane of eigenvalues 
$\Lambda$ for particular values of the parameters
    $\Gamma = 0.025$, $\lambda = 1$ and $\eta_0 = 0.1.$}
    \label{fig_11}
\end{figure}

In the numerical analysis, we rely on equation \eqref{eq-delta}, rewritten as a system of first-order equations in time for the functions $\phi_1 = \delta \phi$ and $\phi_2 = \partial_t \delta \phi$
\begin{equation}
    \begin{gathered}
        \partial_{t'} \phi_1 = \phi_2 ,\\
        \partial_{t'} \phi_2 = \widehat{\mathcal{L}}_1 \phi_1 + \widehat{\mathcal{L}}_2 \phi_2 .
    \end{gathered}
\end{equation}
The linear operators in this equation are as follows
\begin{equation}
    \widehat{\mathcal{L}}_1 = (1-v^2) \partial^2_{x'} + \eta_0 v \partial_{x'} - \lambda ( 3 \phi_0^2 -1) , \qquad \widehat{\mathcal{L}}_2 = 2 v \partial_{x'} - \eta_0 .
\end{equation}
The above system of equations can be written in a compact form
\begin{equation}
    \partial_{t'} \Phi = M \Phi ,
\end{equation}
where we have adopted the following notation
\begin{equation}
\Phi = \begin{pmatrix}
\phi_1 \\
\phi_2
\end{pmatrix},  \qquad M= \begin{pmatrix}
0 & 1 \\
\widehat{\mathcal{L}}_1 & \widehat{\mathcal{L}}_2
\end{pmatrix}.
\end{equation}
After separating the time dependence $\Phi(t',x') = e^{\Lambda t'} \Psi(x')$, we arrive at classical (linear) eigenvalue problem 
\begin{equation}
\label{eigen}
     M \Psi =  \Lambda \Psi.
\end{equation}
The eigenvalues obtained based on equation \eqref{eigen} are shown in Figure~\ref{fig_11}. In the central part of the figure, there are values corresponding to oscillatory modes. The red points represent the discrete mode, while the blue lines correspond to oscillations belonging to the continuous spectrum. The zero mode is represented by a green point on the right side of the figure, whereas the magenta point on the left side corresponds to a damped mode. It is worth noting that all eigenvalues have a negative real part (apart from the zero mode), i.e., they correspond to stable excitations. Although the figure was generated for a particular choice of parameters 
($\Gamma = 0.025$, $\lambda = 1$, $\eta_0 = 0.1$), its overall structure is entirely typical.

\begin{figure}[h!]
    \centering
    \subfloat{{\includegraphics[height=4.5cm]{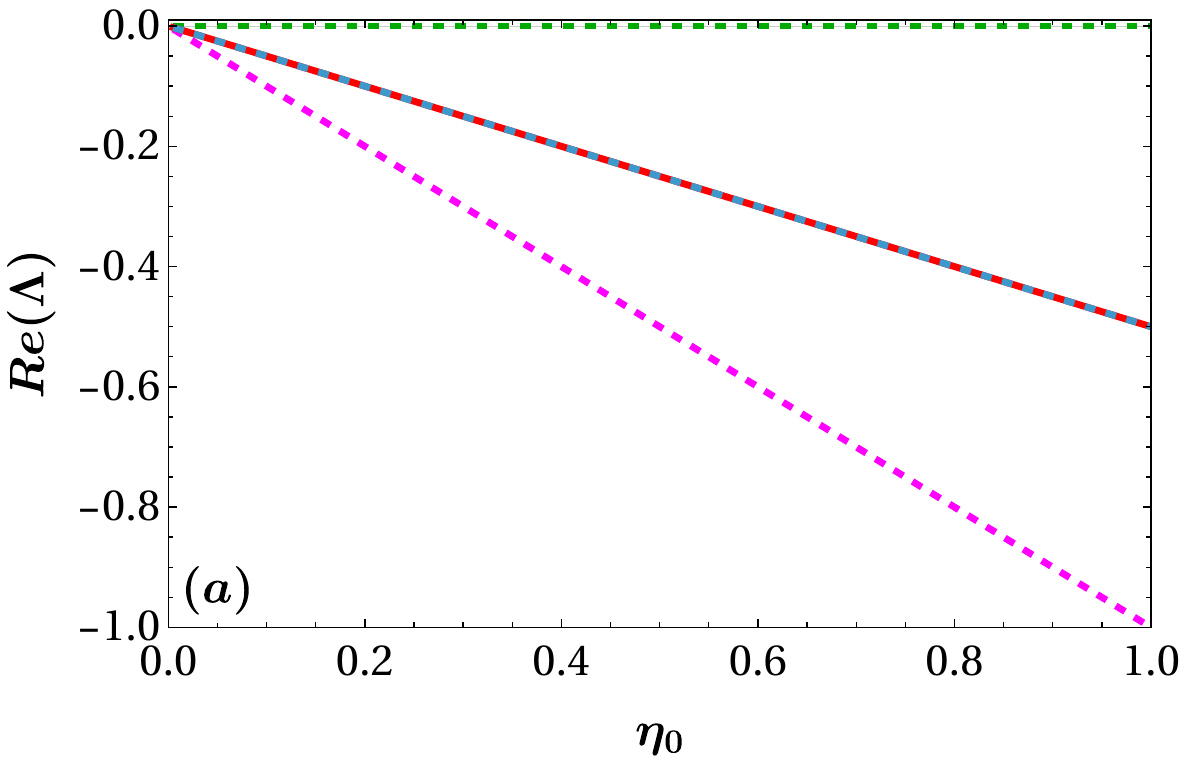}}}
    \quad
    \subfloat{{\includegraphics[height=4.5cm]{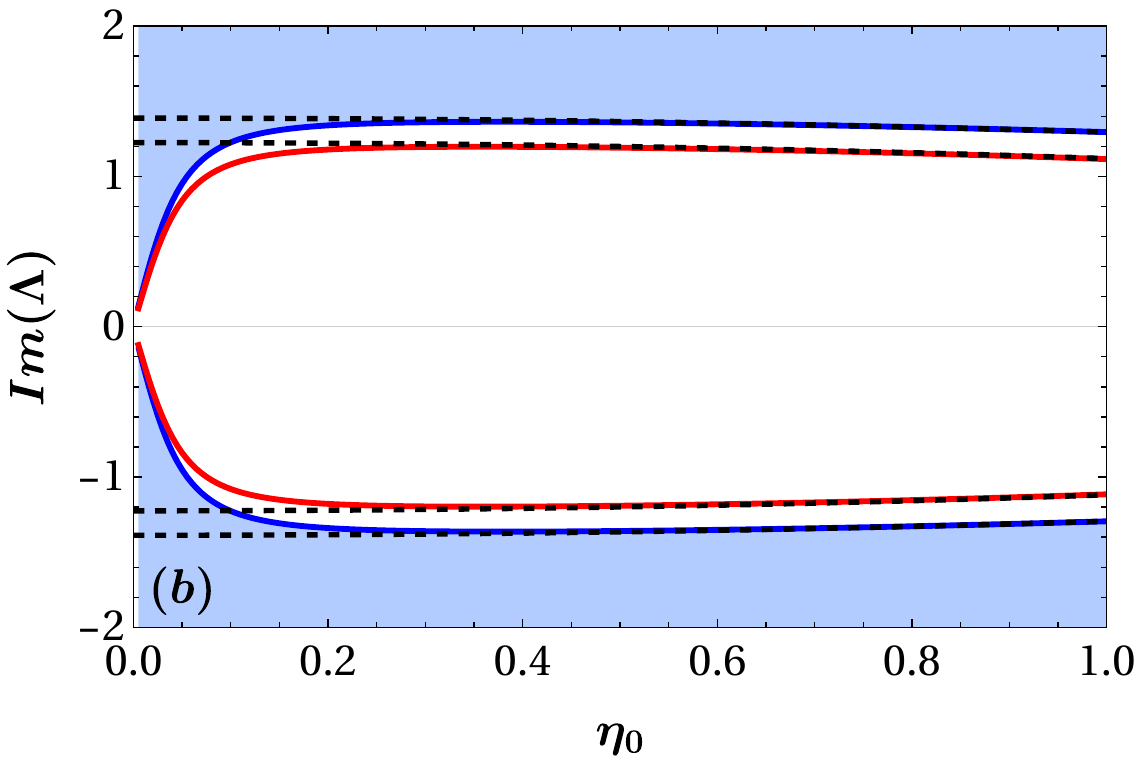}}}
    \caption{The dependence of the eigenvalues of the linear perturbation problem on the damping in the system for
     $\Gamma = 0.025$ and $\lambda = 1$.
     Panel (a) shows the real part of the eigenvalues. The green dashed horizontal line represents the zero mode. The central blue–red line corresponds to the real part of the oscillatory mode and is located at $Re(\Lambda)=-\eta_0/2$. The lower magenta dashed line denotes the purely damped mode with damping coefficient $Re(\Lambda)=-\eta_0$. Panel (b) illustrates dependence of the imaginary part of $\Lambda$ on the damping coefficient $\eta_0$. The blue shaded region represents the continuous spectrum, while the red curve corresponds to the discrete oscillatory mode. The black dashed lines indicate the analytical predictions: the dependence of the discrete mode frequency on $\eta_0$ given by Eq.~\eqref{omega-dis}, and the lower edge of the continuous spectrum determined by Eq.~\eqref{omega-cont}.}
    \label{fig_12}
\end{figure}

Figure~\ref{fig_12} illustrates the variation of the real and imaginary parts of the eigenvalues as a function of the damping coefficient. Panel (a) shows the dependence of the real part of $\Lambda$ on $\eta_0$. The green dashed horizontal line represents the zero mode. The central blue–red line depicts the real part of the oscillatory modes. This line corresponds to the value $Re(\Lambda) =-\eta_0/2$. The lowest magenta dashed line illustrates pure damping with coefficient $Re(\Lambda)=-\eta_0$. Panel (b) shows the dependence of the imaginary part of $\Lambda$ on the coefficient $\eta_0$. The blue region corresponds to the continuous spectrum, while the red line indicates the discrete mode. The imaginary values displayed in this panel are associated with the central blue–red line in panel (a), which represents the real part. As $\eta_0$ approaches zero, the kink velocity increases toward the speed of light. To avoid entering the ultra-relativistic regime, we restricted the dissipation to values not smaller than $\eta_0 = 0.003$, limiting the velocity range from zero up to $v = 0.998$. Moreover, in the figure we have indicated with black dashed lines the analytical results for the dependence of the discrete mode frequency on $\eta_0$ according to formula \eqref{omega-dis}, as well as the lower bound of the continuous spectrum described by formula \eqref{omega-cont}. It can be seen that the agreement between the analytical and numerical results is very good for $\eta_0 > 0.17$, which corresponds to non-relativistic velocities
{i.e., in the present context ones not exceeding $0.3$. This result is not surprising, since the transition to the comoving frame is achieved by means of a Galilean transformation (which is
no longer suitable once the velocities become sufficiently large).}

\begin{figure}[h!]
    \centering
    \subfloat{{\includegraphics[height=4.5cm]{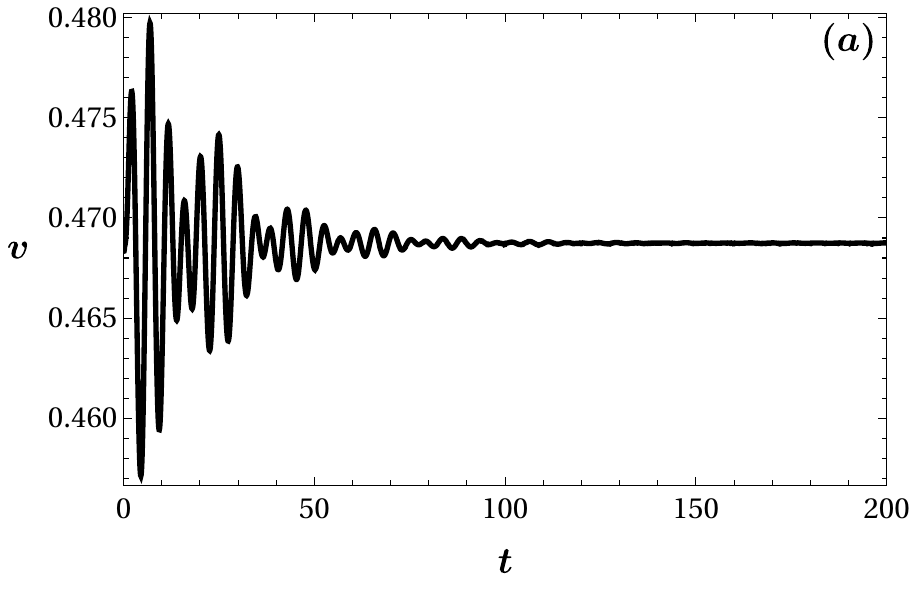}}}
    \quad
    \subfloat{{\includegraphics[height=4.5cm]{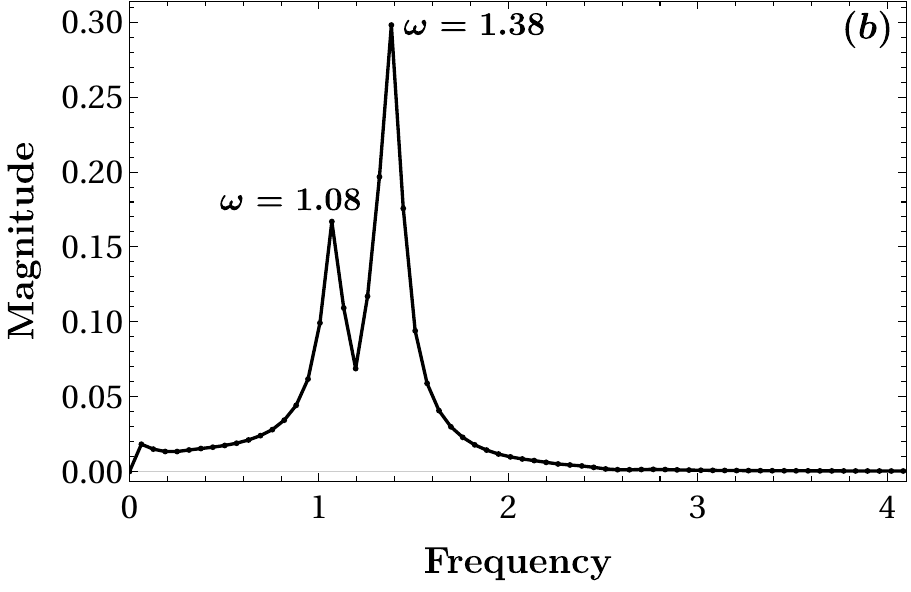}}}
    \caption{
    (a) Velocity variations as a function of time for initial conditions \eqref{phi_wp1-kink}, \eqref{phi_wp2-kink}. In the figure, $\Gamma = 0.025$, $\lambda = 1$, $\eta_0 = 0.1$ was assumed.
(b) Fourier spectrum of the signal shown in panel (a).
 }
    \label{fig_13}
\end{figure}

The above considerations regarding the spectrum of linear excitations of the kink are directly related to the initial oscillations of the kink velocity observed in the simulations presented in this paper.
Strictly speaking, we observe velocity variations at the early stages of the evolution, which manifest themselves in velocity–time plots as damped oscillations. An example of this behavior is shown in Fig.~\ref{fig_13} (a). This plot was obtained for $\lambda=1$, $\Gamma=0.025$, and for a spatially constant dissipation coefficient $\eta_0=0.1$ throughout the system. A Fourier analysis of this signal is presented in panel (b) of the same figure. It turns out that two peaks are present in the spectrum. The peak at the lower frequency, $\omega=1.08$, corresponds exactly to the frequency of the discrete mode from the linear approximation. The second mode, in turn, is located in the continuous part of the spectrum and has a frequency of $\omega=1.38$. This reasoning shows that the oscillations are indeed a consequence of kink excitation caused by a mismatch between the initial configuration and the exact kink solution of the model with external forcing.

\FloatBarrier
\section{Appendix}
\label{AppendixB}
\setcounter{equation}{0}
For the sake of completeness, in this section we present
 results that provide insight into the behavior of traveling wave solutions when the model parameters are simple constants, i.e., $\eta_0=const$ and $\lambda=const$. This homogeneous case serves as a reference point for studying more complex, i.e., heterogeneous, cases. Let us consider a solution in the form of a traveling wave, i.e., $\phi=\phi(z)$ where $z=x -  v t$. In this case, the equation \eqref{phi4+} reduces to an ordinary equation of the form
\begin{equation}
\label{wave-phi4}
 (v^2-1) \frac{d^2 \phi}{d z^2} - v  \eta_0 \frac{d \phi}{d z}+ \lambda \, \phi (\phi^2 -1)  = 0 .
\end{equation}
Multiplying the above equation by the derivative of the function $\phi$ with respect to the variable $z$, and then integrating with respect to this variable, allows us to determine the admissible velocities that a wave traveling in this system can attain
\begin{equation}
\label{wave-phi4+}
(v^2-1) \int_{-\ell}^{+\ell} d z \, \frac{d \phi}{d z} \frac{d^2 \phi}{d z^2} - v  \eta_0 \int_{-\ell}^{+\ell} d z \, \left(\frac{d \phi}{d z}\right)^2 + \lambda \, \int_{-\ell}^{+\ell} d z \,\frac{d \phi}{d z} \,\phi (\phi^2 -1)  = 0 .
\end{equation}
For a half-kink solution we impose the boundary conditions $\phi(-\ell)=1$ and $\phi(+\ell)=0$, which imply that the third integral evaluates to $\frac{1}{4}$. Moreover, since the derivative vanishes at large $\pm\ell$, the first integral is zero. Denoting by $\mathcal{N}$ the integral of the middle term
\begin{equation}
\label{N}
    \mathcal{N} = \int_{-\ell}^{+\ell} dz \left(\frac{d \phi}{d z} \right)^2, 
\end{equation}
we obtain an expression for the velocity of the arbitrary half-kink like configuration in a homogeneous system
\begin{equation}
\label{v-field}
    v=\frac{\lambda}{4  \, \eta_0 \, \mathcal{N}}.
\end{equation}
Next we obtain the exact form of the solution in the form of a half-kink based on the following test function
\begin{equation}
    \label{half-phi}
    \phi(t,x) = \frac{1}{1+e^{\xi(t,x)}} .
\end{equation}
The derivatives of this function with respect to $\xi$ take a simple polynomial form in the field $\phi$ itself
\begin{equation}
    \label{poly}
    \partial_{\xi} \phi(\xi) = - \phi (1 - \phi) \,\,\,\, \mathrm{and} \,\,\,\, \partial^2_{\xi} \phi(\xi) =  \phi (1 - \phi) (1 - 2 \phi).
\end{equation}
Substituting these expressions into equation of motion allows us to derive the differential constraints on an arbitrary function $\xi(t,x)$
\begin{equation}
      \left( - \partial^2_t \xi - \eta_0 \partial_t \xi +  \, \partial^2_x \xi 
       + \left( \partial_t \xi\right)^2 - \left( \partial_x \xi\right)^2
       - \lambda \right)
        - 2 \left(  \left( \partial_t \xi\right)^2 - \left( \partial_x \xi\right)^2
       + \frac{\lambda}{2}\right) \phi =0 .
\end{equation}
By setting the coefficients of the powers of $\phi$ to zero and simplifying the resulting expressions, we obtain a system of differential equations for the function $\xi=\xi(t,x)$
\begin{equation}
    \begin{gathered}
     - \partial^2_t \xi - \eta_0 \partial_t \xi +  \, \partial^2_x \xi 
      = \frac{3 \lambda}{2} ,
       \\  
       \left(\partial_t \xi\right)^2 - \left( \partial_x \xi\right)^2
       = - \frac{\lambda}{2} .
    \end{gathered}
\end{equation}
If $\xi(t,x)$ is linear in the variables, then this amounts to an
algebraic condition. In this case the solution can be written as:
\begin{equation}
\xi(t,x) = 
\sqrt{\frac{\lambda}{2 }}\,\sqrt{1 + \frac{9\lambda}{2\eta_0^{2}}}\, x 
- \frac{3\lambda}{2\eta_0}\, t .
\end{equation}
A more transparent representation of this solution is given by:
\begin{equation}
\label{xi-sol}
\xi(t,x) = 
\sqrt{\frac{\lambda}{2 }}\,\gamma \left( x 
-\,v  t \right) \,\,\,\, \mathrm{where} \,\,\,\, v = \frac{ 1}
   {\sqrt{1+\frac{2 \eta_0^2}{9 \lambda}}} \,\,\,\, \mathrm{and} \,\,\,\, \gamma=\frac{1}{\sqrt{1-v^2}}.
\end{equation}
For the homogeneous case, $\gamma$ can be naturally interpreted as the Lorentz factor.
It is worth noting that if we calculate the integral $\mathcal{N}$ for the solution \eqref{half-phi}, \eqref{xi-sol}, we obtain 
\begin{equation}
    \mathcal{N} =\frac{\sqrt{\lambda} }{12 \eta_0} \sqrt{2 \eta_0^2 + 9 \lambda}
\end{equation}
which, when substituted into equation \eqref{v-field}, reproduces the velocity from equation \eqref{xi-sol}.

\section{Appendix}
\label{AppendixC}
\setcounter{equation}{0}
This appendix defines the functions involved in the coefficients of effective equations.
All integrals are evaluated over the interval $(\xi_-,\xi_+)$ where
\begin{equation}
\xi_{-}=\sqrt{\frac{\lambda}{2}} \gamma_0 \left( -\ell - x_0\right), \,\,\,\,\, \xi_{+}=\sqrt{\frac{\lambda}{2}} \gamma_0 \left( \ell - x_0\right) ,
\end{equation}
for model 1,
\begin{equation}
\xi_{-}=\sqrt{\frac{\lambda}{2}} \gamma \left( -\ell - x_0\right), \,\,\,\,\, \xi_{+}=\sqrt{\frac{\lambda}{2}} \gamma \left( \ell - x_0\right) ,
\end{equation}
for model 2, and
\begin{equation}
\xi_{-}=\sqrt{\frac{\lambda}{2}} g(x_0) \left( -\ell - x_0\right), \,\,\,\,\, \xi_{+}=\sqrt{\frac{\lambda}{2}} g(x_0) \left( \ell - x_0\right) ,
\end{equation}
for model 3.

\begin{figure}[h!]
    \centering
    \subfloat{{\includegraphics[height=4.5cm]{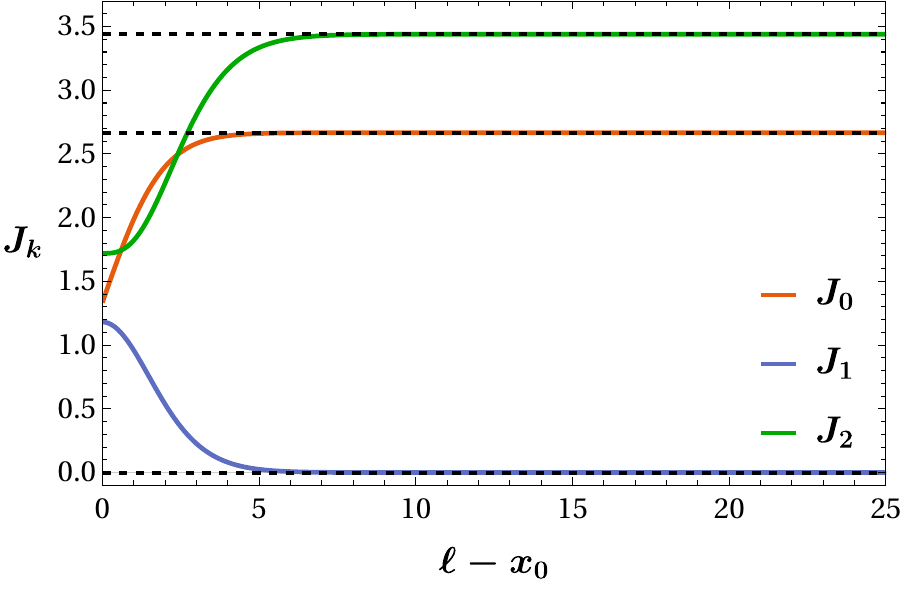}}}
    \caption{Dependence of the integrals $J_k$, where $k = 0,1,2$, on the variable $\ell - x_0$. The dashed black lines correspond to the asymptotic values.}
    \label{fig_14}
\end{figure}

\noindent
The integrals themselves are defined below. The first function has the form
\begin{equation}
\begin{gathered}
    J_0(x_0,\gamma,\ell) = \int_{\xi_{-}}^{\xi_{+}} d \xi \, \, \sech^4 \frac{\xi}{2} = \\  
  \frac{2}{3}  \left[
\bigl(2 + \cosh \xi_{+}\bigr)\, \sech^2 \frac{\xi_{+}}{2} \, \tanh \frac{\xi_{+}}{2}
- \bigl(2 + \cosh \xi_{-}\bigr)\, \sech^2 \frac{\xi_{-}}{2} \, \tanh \frac{\xi_{-}}{2} 
 \right],
  \end{gathered}
  \end{equation}
As $\ell \rightarrow \infty$, this integral simplifies significantly
    \begin{equation}
      \lim_{\ell \rightarrow \infty} J_0 = \frac{8}{3}   .
    \end{equation}
The second integral is as follows
    \begin{equation}
        \begin{gathered}
    J_1(x_0,\gamma,\ell) = \int_{\xi_{-}}^{\xi_{+}} d \xi \, \xi \, \sech^4 \frac{\xi}{2} = 
   \frac{2}{3} \Bigg[ 
4  \ln \left(\frac{\cosh\frac{\xi_{-}}{2}}{\cosh\frac{\xi_{+}}{2}}\right) +
\\
\sech^2 \frac{\xi_{+}}{2} \,
\Bigl(1 + \xi_{+} \bigl(2 + \cosh \xi_{+}\bigr) \tanh \frac{\xi_{+}}{2} \Bigr)
- \sech^2 \frac{\xi_{-}}{2} \,
\Bigl(1 + \xi_{-} \bigl(2 + \cosh \xi_{-}\bigr) \tanh \frac{\xi_{-}}{2} \Bigr) \Bigg] .   
\end{gathered}
\end{equation}
In the limit $\ell \rightarrow \infty$, it disappears
  \begin{equation}
      \lim_{\ell \rightarrow \infty} J_1 = 0  .
    \end{equation}
The value of the last integral is given by the following expression
    \begin{equation}
        \begin{gathered}
    J_2(x_0,\gamma,\ell) = \int_{\xi_{-}}^{\xi_{+}} d \xi \, \xi^2 \, \sech^4 \frac{\xi}{2} = \\
\frac{4}{3} \Bigg[\xi_{-}^2 - \xi_{+}^2
+ 4 \, \xi_{-} \, \ln\!\bigl(1 + e^{-\xi_{-}}\bigr)
- 4 \, \xi_{+} \, \ln\!\bigl(1 + e^{-\xi_{+}}\bigr)
- 4 \, \operatorname{Li}_2\!\bigl(-e^{-\xi_{-}}\bigr)
+ 4 \, \operatorname{Li}_2\!\bigl(-e^{-\xi_{+}}\bigr) +
\\
\frac{(-2 + \xi_{+}^2)\sinh \xi_{+} + \xi_{+}\bigl(2 + \xi_{+} \tanh\frac{\xi_{+}}{2}\bigr)}{1 + \cosh \xi_{+}} 
-
\frac{(-2 + \xi_{-}^2)\sinh \xi_{-} + \xi_{-}\bigl(2 + \xi_{-} \tanh\frac{\xi_{-}}{2}\bigr)}{1 + \cosh \xi_{-}} 
\Bigg] .
\end{gathered}  
\end{equation}
In the limit $\ell \to \infty$, this expression is significantly simplified
  \begin{equation}
      \lim_{\ell \rightarrow \infty} J_2 = 8 \left(\frac{\pi^2}{9} -\frac{2}{3}  \right) .
    \end{equation}
It should be noted that deviations from the asymptotic values occur only for relatively small systems, which lie outside the scope of this study. {Figure~\ref{fig_14} shows how quickly the asymptotic regimes are reached by the individual integrals appearing in the paper. The orange line shows the dependence of $J_0$ on $\ell - x_0$, the green line shows $J_1$, and the blue line shows $J_2$. The dashed black lines indicate the asymptotic values of these integrals. } 

\section{Appendix}
\label{AppendixD}
\setcounter{equation}{0}
We use the function $\tilde{g}(x_0)= \frac{1}{\gamma_0} \,g(x_0)$ from equation \eqref{xi-n}, which defines the ansatz. The function contains five coefficients $a_i$, but they are  functions of $\varepsilon$ and $\eta_0$. In fact, the function $\Tilde{g}$ contains no free parameters other than those already present in the underlying field model, and the initial position of the half-kink
\begin{equation}
\begin{gathered}
\label{tilde-g1}
    \Tilde{g}(x_0;\eta_0,\varepsilon) = 1 + 2(a_1+a_2x_0+a_3x_0^2)\left[\frac{1}{1+e^{\frac{1}{\sqrt{2}} (x_0-\ell)}} - \frac{1}{1+e^{\frac{1}{\sqrt{2}} (x_0)}}\right]\\-(a_4+a_5x_0)\left[\sech^2\left(\frac{1}{2\sqrt{2}}(x_0-\ell)\right)-\sech^2\left(\frac{1}{2\sqrt{2}}x_0\right)\right].
\end{gathered}
\end{equation}
Note that the first square bracket contains the difference between the values of the function describing the half-kink located in the first and the second medium, while the second square bracket contains the difference between their derivatives.
The dependence of $a_i$ on the value of the dissipation coefficient $\eta_0$ and on the value of the change in dissipation $\varepsilon$ when transitioning from one medium to another is the following
\begin{equation}
    \begin{aligned}
        \label{ai-1}
a_1(\eta_0,\varepsilon) &= 10^{-2}\Big[-2.26 (1-1.27 \eta_0+0.179 \eta_0^2-0.00747 \eta_0^3) \varepsilon\\&-4.64 (1-0.123 \eta_0+0.00301 \eta_0^2) \varepsilon^2+1.1 (1-0.113 \eta_0) \varepsilon^3\Big],\\
        a_2(\eta_0,\varepsilon) &=  10^{-4}\Big[-5.74 (1 + 0.634 \eta_0 - 0.0972 \eta_0^2 + 0.00370 \eta_0^3) \varepsilon\\&+ 17.7 (1 - 0.0975 \eta_0) \varepsilon^2 - 5.96 (1 - 0.116 \eta_0) \varepsilon^3\Big],\\
        a_3(\eta_0,\varepsilon) &= 10^{-6}\Big[2.67(1 + 0.905 \eta_0 - 0.121 \eta_0^2 + 0.00396 \eta_0^3) \varepsilon\\&- 14.9 (1 - 0.0993 \eta_0) \varepsilon^2 + 5.92(1 - 0.119 \eta_0) \varepsilon^3\Big],\\
        a_4(\eta_0,\varepsilon) &= 1.3\cdot10^{-4}(1 + 307 \varepsilon - 39.5 \varepsilon^2)(1-0.136 \eta_0 + 0.00620 \eta_0^2),\\
        a_5(\eta_0,\varepsilon) &= 1.84\cdot10^{-3}(1 + 127 \varepsilon + \varepsilon^2)(1 - 0.132 \eta_0 + 0.00593 \eta_0^2).
    \end{aligned}
\end{equation}
The above function allows for a very precise determination of the changes in the half-kink velocity during its propagation through the system in a wide range of parameters $\eta_0$ and $\varepsilon$.

\noindent
On the other hand the rescaled function $g(x_0)$, i.e., $\tilde{g}(x_0) = \frac{1}{\gamma_0} g(x_0)$, defining the ansatz \eqref{xi-n} in the case of a layer with fixed thickness contains eight coefficients 
\begin{equation}
\begin{gathered}
\label{tilde-g-2}
\Tilde{g}    (x_0;\eta_0,\varepsilon) = 1 + 2(a_1+a_2x_0+a_3x_0^2)\left[\frac{1}{1+e^{\frac{1}{\sqrt{2}} (x_0-\mathtt{L}_0)}} - \frac{1}{1+e^{\frac{1}{\sqrt{2}} (x_0)}}\right]\\-a_4\left[\sech^2\left(\frac{1}{2\sqrt{2}}(x_0-\mathtt{L}_0)\right)-\sech^2\left(\frac{1}{2\sqrt{2}}x_0\right)\right]\\+2\left(a_5+a_6 x_0+a_7 x_0^2\right)\left[\frac{1}{1+e^{\frac{1}{\sqrt{2}} (x_0-\mathtt{L}_0)}}+1\right]\sech^2(a_8 (x_0-\mathtt{L}_0)).
\end{gathered}
\end{equation}
This quantity is constructed from functions $a_i = a_i(\eta_0, \varepsilon)$ which for $\mathtt{L}_0 = 40$ have the form
\begin{equation}
    \begin{aligned}
        \label{ai-2}a_1(\eta_0,\varepsilon) &= 2.68\cdot10^{-2}(1+0.255 \eta_0-0.0219 \eta_0^2)(1-0.238 \varepsilon)\varepsilon,\\
        a_2(\eta_0,\varepsilon) &= -3.03\cdot10^{-3}(1-0.0527 \eta_0)(1-0.192 \varepsilon)\varepsilon,\\
        a_3(\eta_0,\varepsilon) &= 5.12\cdot10^{-5}(1-0.0649 \eta_0) (1-0.0580 \varepsilon)\varepsilon,\\
        a_4(\eta_0,\varepsilon) &= 2.35\cdot10^{-1}(1-0.134 \eta_0+0.00602 \eta_0^2)(1-0.0674 \varepsilon)\varepsilon,\\
        a_5(\eta_0,\varepsilon) &= 1.5\cdot10^{-1}(1 + 0.0131 \eta_0 - 0.0076 \eta_0^2 + 0.000284 \eta_0^3) (1 + 0.554 \varepsilon - 0.487 \varepsilon^2)\varepsilon,\\
        a_6(\eta_0,\varepsilon) &= 4.43\cdot10^{-3}(1- 0.0406 \eta_0 - 0.000322 \eta_0^2)(1 + 0.925 \varepsilon - 0.672 \varepsilon^2)\varepsilon,\\
        a_7(\eta_0,\varepsilon) &= -4.51\cdot10^{-5}(1 - 0.0664 \eta_0 + 0.0013 \eta_0^2)(1 + 1.17 \varepsilon - 0.637 \varepsilon^2)\varepsilon,\\
        a_8(\eta_0,\varepsilon) &= -1.14\cdot10^{-1}(1 + 0.216 \eta_0 - 0.0229 \eta_0^2 + 0.000831 \eta_0^3) (1 + 0.319 \varepsilon - 0.0904 \varepsilon^2 + 0.00697 \varepsilon^3).
    \end{aligned}
\end{equation}
 As in the previous case, the agreement with the results obtained from the field model is  remarkably good.
 
\FloatBarrier
\printbibliography

\end{document}